\documentclass[sigconf]{acmart}

\usepackage{booktabs} 
\usepackage{multirow} 
\usepackage{adjustbox} 
\usepackage{siunitx}
\usepackage{array}  
\usepackage{tabularx}
\usepackage{caption}
\usepackage{colortbl}
\usepackage{xcolor}
\usepackage{makecell}
\usepackage{libertine}
\usepackage{xspace}
\usepackage{arydshln}
\usepackage{enumerate}
\usepackage{enumitem}
\usepackage{pifont}

\usepackage{tikz}

\newcommand{\blackcircled}[1]{%
  \tikz[baseline=(char.base)]{%
    \node[shape=circle, fill=black, inner sep=1pt, text=white] (char) {#1};%
  }%
}
\newcommand{\DomainCodeBench}{\textsc{DomainCodeBench}\xspace}%
\usepackage[most]{tcolorbox}
\newcommand{\figmargin}{\vspace{-4pt}}
\newcommand{\tabmargin}{\vspace{-4pt}}

\newcommand{\boxmargin}{2mm}
\tcbuselibrary{most, skins, breakable, theorems}
\newtcolorbox{myboxa}[2][]{
    colback=gray!10!white,
    colframe=black, enhanced,
    attach boxed title to top left={yshift=-2mm,xshift=5mm},
    title=#2,#1
}
\newtcolorbox{myboxb}[2][]{
    boxsep=1pt,
    left = \boxmargin, right = \boxmargin, top = \boxmargin, bottom = \boxmargin,
    title={#2},#1
}

\newtcolorbox{myboxc}{
    colback=yellow!10!white,    
    colframe=gray!50,           
    arc = 0pt, outer arc = 5pt,
    boxsep=0pt, left = 3pt, right = 0pt, top = 0pt, bottom = 0pt, 
    leftrule=3pt,               
    bottomrule=0pt, toprule=0pt, rightrule=0pt,
    left = \boxmargin, right = \boxmargin, top = \boxmargin, bottom = \boxmargin,
    before skip=5pt,   
    after skip=5pt    
}

\begin{document}
\title{Benchmarking LLMs' Code Generation Capacity in Multi Read-world Software Application Domains}
\title{Can LLMs Generate Code for Different Application Domains? - An Experimental Evaluation}
\title{Can LLMs Generate Code for Different Application Domains? - Benchmark and Evaluation}
\title{How Well Do LLMs Generate Code for Different Application Domains? Benchmarking and Evaluation}
\title{How Well Do LLMs Generate Code for Different Application Domains? Benchmark and Evaluation}






\title{Top General Performance = Top Domain Performance? \textsc{DomainCodeBench}: A Multi-domain Code Generation Benchmark}



\author{Dewu Zheng}
\affiliation{
\institution{Sun Yat-sen University}
\country{}
}
\email{zhengdw5@mail2.sysu.edu.cn}

\author{Yanlin Wang}
\affiliation{
\institution{Sun Yat-sen University}
\country{}
}
\email{wangylin36@mail.sysu.edu.cn}

\author{Ensheng Shi}
\affiliation{
\institution{Huawei Cloud Computing Technologies Co., Ltd}
\country{}
}
\email{shiensheng@huawei.com}

\author{Xilin Liu}
\affiliation{
\institution{Huawei Cloud Computing Technologies Co., Ltd}
\country{}
}
\email{liuxilin3@huawei.com}

\author{Yuchi Ma}
\affiliation{
\institution{Huawei Cloud Computing Technologies Co., Ltd}
\country{}
}
\email{mayuchi1@huawei.com}

\author{Hongyu Zhang}
\affiliation{
\institution{Chongqing University}
\country{}
}
\email{hyzhang@cqu.edu.cn}

\author{Zibin Zheng}
\affiliation{\institution{Sun Yat-sen University}
\country{}
}
\email{zhzibin@mail.sysu.edu.cn}

\settopmatter{printacmref=false}
\setcopyright{none}
\renewcommand\footnotetextcopyrightpermission[1]{}

\begin{abstract}
    
With the rapid advancement of large language models (LLMs), extensive research has been conducted to investigate the code generation capabilities of LLMs. 
However, existing efforts primarily focus on general-domain tasks, leaving LLMs' code generation performance in real-world application domains underexplored. 
This raises a critical question: can a model's general-domain coding ability reliably represent its ability in specialized domains?
In this paper, we introduce \DomainCodeBench, a multi-domain code generation benchmark designed to systematically evaluate LLMs across 12 software application domains and 15 programming languages.
\DomainCodeBench contains 2,400 manually verified tasks with ground-truth, human-annotated docstrings, and fine-grained dependency information to ensure more coverage of domain-specific challenges. 
Specifically, we first identify the most popular application domains by topic mining. Then, we curate coding tasks based on commonly used frameworks and platforms in each domain. 
We obtain several findings through extensive experiments on \DomainCodeBench with ten mainstream LLMs. 
(1) \textbf{Performance decoupling:} experiments reveal that top general-domain models do not consistently excel in specific application domains; 
(2) \textbf{Domain-specific weaknesses:} LLMs often fail due to domain knowledge gaps and third-party library misusage;
(3) \textbf{Contextual enhancement:} we show that augmenting prompts with domain-specific knowledge improves performance by around 38.17\%, providing actionable insights for performance optimization.
Our replication package, including the benchmark, source code, and experimental results, is available at \url{https://github.com/DeepSoftwareAnalytics/DomainCodeBench}. 

\end{abstract}


\maketitle

\section{Introduction}
Large language models have demonstrated remarkable capabilities in code generation and are being rapidly applied in real-world software development scenarios~\cite{survey1,survey2,dong2024self,jiang2023selfevolve,githubcopilot,metagpt,chatdev,codegeneration1,codegeneration2,le2022coderl,Cursor,li2024enhancing,pang2024ai2apps,tao2024magis,wang2024beyond,guo2024stop}. With the rapid increase in the number of LLMs, many benchmarks have been proposed to evaluate the code generation capability of LLMs~\cite{humaneval,naturalcodebench,aixbench,multipl,mbpp,classeval,evoeval,repocoder,codereval,crosscodeeval,evocodebench,repositorylevel,repobench}. 
For example, HumanEval~\cite{humaneval}, a manually designed benchmark, is commonly used to assess LLMs' fundamental programming abilities. Recently, several repository-level code generation benchmarks have been proposed to assess LLMs' code generation capability in real development scenarios, such as CoderEval~\cite{codereval}, EvoCodeBench~\cite{evocodebench}, and HumanEvo~\cite{humanevo}.

However, existing code generation benchmarks primarily focus on general-domain code generation capabilities of LLMs (e.g., generic algorithms and mathematical problems) while leaving assessing their performance in specific application domains largely underexplored~\cite{codegen4libs,private,gu2024effectiveness,tang2024biocoder,zan2022cert}. Compared to general domains, specific application domains may impose unique challenges on LLMs when performing code generation tasks as they may involve distinct programming languages, specific requirements, numerous third-party libraries, and diverse development platforms~\cite{ryder2005impact,croft2022empirical}.
This raises a critical question: \textbf{\textit{can a model's general-domain coding ability reliably represent its ability in specialized application domains?}}

\begin{table*}[t]
\centering
\small
\caption{Comparison of existing benchmarks and \DomainCodeBench. \#Task stands for the number of task instances included; \#Token stands for the average number of tokens per instance.}
\label{table:Comparison-between-existing-popular-benchmarks-and-DomainCodeBench.}
\tabmargin
\begin{tabular}{|l|cc|>{\centering\arraybackslash}m{7cm} |>{\centering\arraybackslash}m{4cm}| } 
\hline
\textbf{Benchmark} & \textbf{\#Task} & \textbf{\#Token} & \textbf{Domain} & \textbf{Language} \\
\hline
HumanEval~\cite{humaneval} & 164 & 57.57 & General Domains & Python \\ 
MBPP~\cite{mbpp} & 974 & 48.6 & General Domains & Python \\  
NumpyEval~\cite{zan2022cert} & 101 & 29.9 & Public Library & Python \\ 
TorchDataEval~\cite{zan-etal-2022-language} & 50 & 50.7 & Private Library & Python \\ 
CoderEval~\cite{codereval} & 460 & 92.84 & General Domains & Python, Java \\ 
DomainEval~\cite{domaineval} & 2454 & 154.07 & Basic Programming Domains 
& Python\\ \hline
\DomainCodeBench & 2400 & 194.2 & Blockchain, Cloud service, Data analysis, Deep learning, Desktop application, Distributed system, Enterprise application, Game, IoT, Mobile, Robotics, Web 
& Python, JavaScript, TypeScript, Solidity, Go, C\# CPP, C, Java, Rust, Scala, PHP, Lua, Kotlin, Swift \\
\hline
\end{tabular}
\end{table*}

\textbf{DomainCodeBench.} In this paper, we introduce a new \emph{multi-domain, human-verified, dependency-enriched} benchmark, \DomainCodeBench, to systematically evaluate LLMs' repository-level code generation capability across 12 prevalent software application domains. 
As shown in Table~\ref{table:Comparison-between-existing-popular-benchmarks-and-DomainCodeBench.}, unlike previous benchmarks that either focus on general domains or isolated technical domains, \DomainCodeBench has broad domain coverage and a large number of task instances. Through large-scale topic mining of technical discussions across Stack Overflow, GitHub, and developer forums, we identify 12 prevalent application domains (e.g., blockchain, robotics), each accompanied by their dominant programming frameworks (e.g., PyTorch for deep learning). Note that DomainEval~\cite{domaineval} is the closest work to our benchmark. Differently, DomainEval focuses on programming domains (e.g., computing, network, visualization) instead of application domains.

\textbf{Construction Process.} The quality of \DomainCodeBench is ensured through a rigorous data construction process. 
Given the vast number of application domains, covering all of them is nearly impossible. We start with conducting a survey of the most frequently discussed application domains between January 1, 2020 and June 1, 2024 to identify the target domains for \DomainCodeBench. 
Next, we sample programming tasks from high-quality real-world projects related to these domains and perform extensive manual annotation to equip each task instance with a high-quality docstring. 
Finally, the resulting \DomainCodeBench comprises 2,400 manually verified task instances, with 200 for each domain.

\textbf{Evaluation \& Findings.} We evaluate the code generation capabilities of 10 mainstream LLMs (including both open-source and closed-source ones) on \DomainCodeBench. 
Experimental results yield the following key findings:
\blackcircled{1} \emph{Performance decoupling.} (i) Strong performance on general-domain benchmarks does not necessarily imply strong performance across all application domains. (ii) Mainstream LLMs still have significant room for improvement in code generation tasks within specific domains. (iii) LLMs exhibit varying levels of performance across different domains, and generally generally perform well in blockchain application development but show weaker performance in domains such as web development. 
\blackcircled{2} \emph{Domain-specific weakness.} We identify several common factors that cause LLMs to fail to generate domain-specific code, including lack of relevant domain knowledge, absence of repository context, and challenges in handling third-party libraries.
\blackcircled{3} \emph{Contextual enhancement.} Domain-specific context can significantly help improve the LLMs' code generation performance in specific domains by around 38.17\% on average.

We summarize the main contributions as follows:
\begin{itemize}[leftmargin=5pt]
    \item We introduce \DomainCodeBench, a new code generation benchmark that spans 12 specific application domains and 15 programming languages.
    \item Through extensive experiments, we reveal the code generation performance of mainstream LLMs in these specific application domains, providing actionable insights for developers and researchers. 
    \item Through in-depth error analysis, we summarize the reasons behind LLMs' suboptimal performance in code generation tasks for specific domains, offering directions for further improving LLMs' capabilities.
    \item We conduct thorough experiments to explore the possibility of improving LLMs' domain-specific code generation capabilities, providing promising directions for further enhancing LLMs' performance in various domains.
    
\end{itemize}

\section{Related Work}

\paragraph{General-domain Code Genetation Benchmarks.} 
To evaluate the code generation capabilities of LLMs, researchers have introduced numerous benchmarks tailored to code generation task~\cite{humaneval,mbpp,aixbench,multipl,classeval,athiwaratkun2022multi,cobbe2021training,humanevo,wang2024sparsecoder}. HumanEval~\cite{humaneval} includes 164 problems constructed manually, which requires LLMs to generate Python function based on NL descriptions. MBPP~\cite{mbpp} contains 974 Python programming problems, covering a range of common tasks such as list processing, string manipulation, and mathematical computations. Although these benchmarks effectively reveal the general-domain code generation capabilities of current LLMs, they overlook the fact that these LLMs will ultimately be applied to a wide range of downstream domains, and performance in general-domain code generation tasks does not necessarily translate to performance in domain-specific development scenarios. Most existing code generation benchmarks primarily assess LLMs' ability to generate general-domain functions without targeting specific domains, which limits their practicality for developers working in specialized software application domains.

\paragraph{Domain-specific Code Generation Benchmarks.} 
Recent efforts have started to shift the focus of code generation benchmarks toward various domains~\cite{domaineval,domcoder,zhuo2024bigcodebenchbenchmarkingcodegeneration}. BigCodeBench~\cite{zhuo2024bigcodebenchbenchmarkingcodegeneration} offers tasks involving multiple function calls, covering 139 libraries and 1,140 tasks across seven fundamental programming domains. 
DomainEval~\cite{domaineval} evaluates LLMs' fundamental programming abilities across multiple domains (e.g., computing, systems, and cryptography). 

However, despite the advancements made by these benchmarks in addressing the need for domain-specific code generation benchmarks, the domains emphasized in these benchmarks still revolve around foundational programming domains rather than the actual application development domains. As a result, developers in downstream application development domains may struggle to directly determine the LLMs' code generation performance in their respective domains based on these benchmarks.

In summary, existing code generation benchmarks primarily focus on evaluating the code generation performance of LLMs in general domains, overlooking their capabilities in specific downstream domains. Although some research has started exploring LLMs' code generation performance across various domains, it mainly focuses on fundamental programming domains rather than real-world application development domains, making these benchmarks unable to provide direct guidance for developers.

\begin{figure}[t] 
    \centering
    \includegraphics[width=0.47\textwidth]{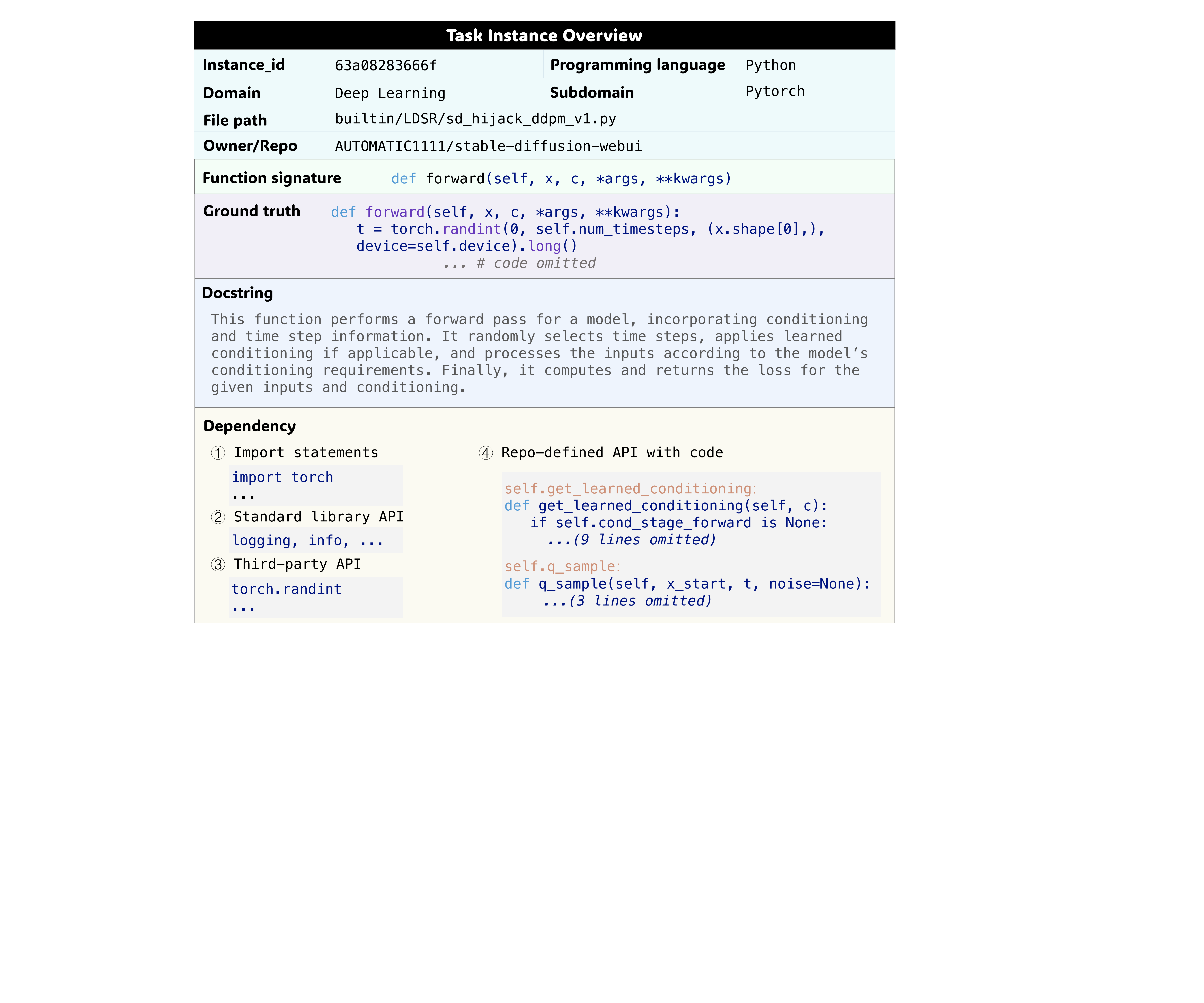} 
    \figmargin
    \caption{An example task instance in \DomainCodeBench.} 
    \label{fig:overview} 
    \Description{}
    
\end{figure}

\begin{table*}[t]
    \centering
    \small   
    \caption{The application domains covered by \DomainCodeBench.}
    \label{table:Studied-detailed-taxonomy-of-domains-covered-by-DomainCodeBench.}
    \tabmargin
    \begin{tabular}{|l |>{\centering\arraybackslash}m{6cm} |>{\centering\arraybackslash}m{7.5cm}|} 
        \hline
        \textbf{Domain} & \textbf{Subdomain} & \textbf{Programming Languages} \\
        \hline
        Block chain & Bitcoin, EOS, Ethereum & JavaScript, TypeScript, C\#, CPP, Python, C, Java, Solidity, Go \\
        \hline
        Web & Django, Vue, Next, Angular, Node, jQuery, React & TypeScript, JavaScript, Python \\
        \hline
        Mobile & IOS, Android & JavaScript, TypeScript, C\#, C, Java, Swift, Kotlin \\
        \hline
        Data analysis & dask, statsmodels, matplotlib, numpy, scikit-learn, pandas, seaborn & Python \\
        \hline
        IoT & Arduino, google iot core, iot, aws iot core & JavaScript, CPP, C, Python, Java, Go \\
        \hline
        Robotics & gazebo, ROS & C, Python, CPP \\
        \hline
        Distributed system & Netflix oos, kafka, ZooKeeper, hdfs & C\#, CPP, Python, Java, Scala \\
        \hline
        Deep learning & Pytorch, Tensorflow & Python \\
        \hline
        Game & Godot Engine, cocos2d-x, Unity3D, phaser, libgdx, Unreal Engine & JavaScript, Lua, C\#, CPP, Java, Kotlin \\
        \hline
        Cloud service & Azure, GCP, AWS & JavaScript, TypeScript, C\#, Python, Go \\
        \hline
        Enterprise application & CMS, CRM, ERP, HRMS, SCM & JavaScript, TypeScript, Lua, Python, Java, Go, PHP \\
        \hline
        Desktop application & GTK, Qt, WinForm, Electron, WPF & JavaScript, TypeScript, C\#, CPP, Rust \\
        \hline
    \end{tabular}
\end{table*}

\section{DomainCodeBench}

\subsection{Benchmark Overview}

\DomainCodeBench consists of 2,400 task instances spanning 12 prevalent application domains and 15 programming languages, with 200 meticulously curated tasks per domain.
Figure~\ref{fig:overview} illustrates an example task instance. Each task is assigned with a unique identifier (\texttt{Instance\_id}) and categorized by its \texttt{Domain} and \texttt{Subdomain} to denote its specific application context. 
The \texttt{Owner/Repo} field specifies the GitHub project from which the task originates. The \texttt{Function signature} field is included as part of the prompt provided to the LLM, specifying input and output of the target function to guide code generation.

Furthermore, each instance in \DomainCodeBench is equipped with rich dependency-related information, such as \texttt{Import statements} and target function's dependency information. \texttt{Import statements} extracted from the local file can help LLMs understand the dependencies between the local file and other parts of the project. Target function's dependency includes \texttt{Standard library API}, \texttt{Repo-defined API with code} and \texttt{Third-party API} (For repo-defined APIs, we extract the corresponding function implementation code to further assist the LLM in understanding the project context.). 

\begin{figure*}[t]
    \centering
    \includegraphics[width=1\textwidth]{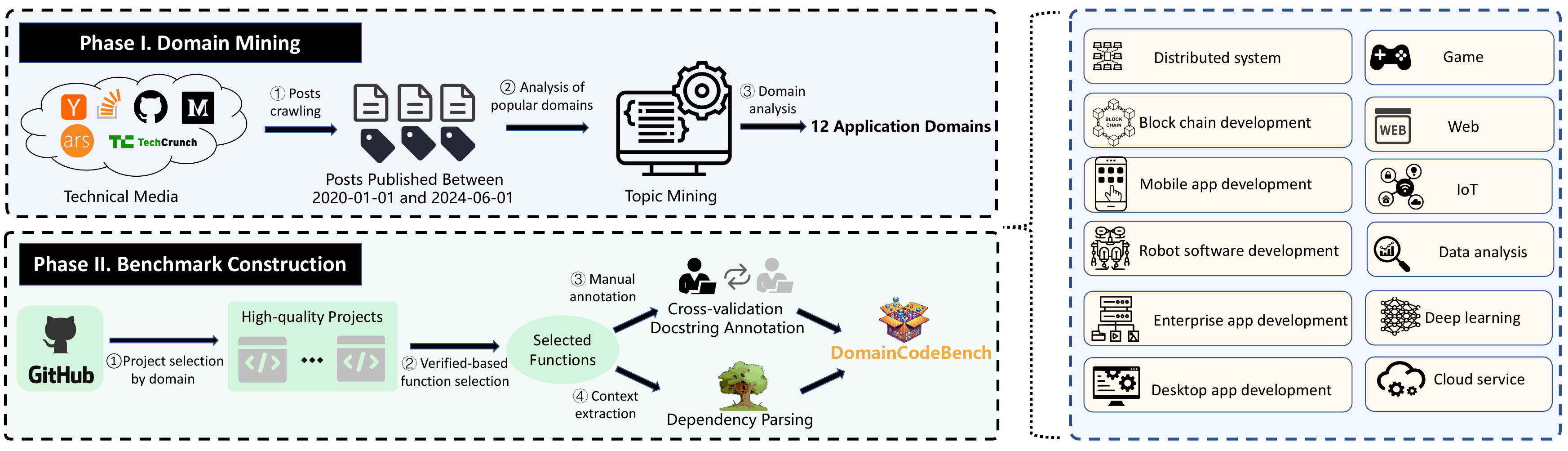} 
    \figmargin
    \figmargin
    \figmargin
    \caption{\DomainCodeBench construction pipeline.}
    \label{fig:construction_pipeline} 
\vspace{-10pt}
\end{figure*}

\subsection{Benchmark Construction Pipeline}

As depicted in Figure~\ref{fig:construction_pipeline}, the construction pipeline of \DomainCodeBench consists of two main phases: domain mining and construction of \DomainCodeBench. 

\subsubsection{Domain Mining.} Due to the large number of software application domains, it is almost impossible to cover all of them. Therefore, we focus on specific domains that are frequently discussed online as target domains for \DomainCodeBench. 

First, we conduct an investigation by analyzing domains with high exposure and frequent discussions between January 1, 2020 and June 1, 2024, on relevant technical websites. Specifically, we collect all posts from active technology news websites (Ars Technica~\cite{arstechnica}, TechCrunch~\cite{TechCrunch}, Hacker News~\cite{HackerNews}), developer communities (Dev~\cite{Dev}, Stack Overflow~\cite{SO}), and blogs (GitHub Blog~\cite{githubblog}, Medium~\cite{medium}) between January 1, 2020, and June 1, 2024. After aggregating all the content, we apply the Latent Dirichlet Allocation (LDA)~\cite{jelodar2019latent} algorithm to identify the underlying topics of the collected posts to identify the most popular domains in the past few years. 
LDA is a topic modeling algorithm that can be used to extract the keywords that represent the underlying topics in NL texts. 
Then, we rank the keywords based on their frequency of occurrence and mannually annotate the domains corresponding to the top 200 keywords. 
Finally, we identify 12 popular application domains that are currently of high interest, such as blockchain, data analysis, etc. All the specific domains studied in \DomainCodeBench are shown in Table~\ref{table:Comparison-between-existing-popular-benchmarks-and-DomainCodeBench.}.

Moreover, we investigate the common technical frameworks and development platforms for each studied domain to more comprehensively evaluate the domain-specific code generation performance of LLMs. As software systems grow in scale and complexity, a single application domain is often divided into several specialized subdomains. Different subdomains within the same domain may rely on different technology stacks, tools, or development platforms, posing distinct challenges to developers. For instance, web development domain can be split into front-end and back-end development, yet the development frameworks they rely on differ significantly. Therefore, to comprehensively reflect the code generation capabilities of LLMs in a specific application domain, we conduct an investigation of the common subdomains within the 12 identified specific domains by analyzing the typical technology frameworks and development platforms used in each subdomain. Table~\ref{table:Studied-detailed-taxonomy-of-domains-covered-by-DomainCodeBench.} shows detailed information on all the covered subdomains in \DomainCodeBench. By starting with this detailed categorization, \DomainCodeBench is designed to offer a thorough evaluation of current LLMs' code generation performance across various application domains. 

\subsubsection{Construction of \DomainCodeBench.} The construction of the \DomainCodeBench benchmark mainly involves four steps, namely, selecting high-quality open-source repositories for specific domains, sampling programming problems from these repositories, manually annotating docstrings through cross-validation, and dependency analysis.

\textbf{(1) Selection of Domain-specific High-quality Projects.} According to the taxonomy in Table~\ref{table:Studied-detailed-taxonomy-of-domains-covered-by-DomainCodeBench.}, we start by searching for real open-source projects on GitHub that are built upon specific technology frameworks or development platforms corresponding to each subdomain. We rank these projects based on their star count, selecting a set of high-quality projects for each domain. The advantage of sampling problems from these high-quality projects is that they generally adhere to good maintenance practices and coding standards, making the selected problems more representative of real-world scenarios.

\textbf{(2) Sampling Programming Problems From the Selected Repositories.} We do not randomly select functions from the selected projects as programming problems. Instead, we conduct careful manual verification to ensure that the chosen programming problems are highly domain-specific rather than general-domain functions. For example, in web development projects, general algorithms and data structure implementations are not considered. Instead, we select functions that are closely related to business logic, including but not limited to core application functionality and functions that invoke domain-specific APIs. Ultimately, we carefully select 2,400 programming problems for \DomainCodeBench, with 200 for each domain.

\textbf{(3) Manually Annotating Docstrings Through Cross Validation.} Since the quality of docstrings in the selected projects can vary and using existing docstrings directly may lead to data leakage issues, we manually rewrite the docstrings for every programming problem in \DomainCodeBench. We conduct a cross-validation-based docstring annotation process involving five annotators, each with more than five years of programming experience. Specifically, each task instance in \DomainCodeBench is assigned to two annotators. An annotator is responsible for writing the requirement description, while the other is tasked with reviewing its correctness and completeness. In cases where the two annotators disagree on the content, a third annotator will be introduced to mediate the discussion. The final version of the docstring will be determined following this arbitration process.
This process ensures that the docstrings in \DomainCodeBench accurately reflect the programming requirements and prevent data leakage by ensuring that LLMs do not directly associate the problems with previously seen training data.

\textbf{(4) Dependency Analysis.} To further assess LLMs' performance in domain-specific code generation, we develop an automated static analysis tool to extract dependencies for the ground truth, which includes standard library APIs, repository-defined APIs, and third-party library APIs used in the ground truth. Since LLMs may not be familiar with repository-defined APIs, we also include the corresponding implementation code for each repository-defined API to help the LLMs better understand the dependencies relevant to the target function.

Finally, we introduce the novel multi-domain code generation benchmark, which contains 2,400 manually verified task instances covering 12 popular application development domains. Through these efforts, we strive to establish \DomainCodeBench as a benchmark for comprehensively evaluating the code generation performance of LLMs across diverse downstream application domains. And we are planning to expand its coverage to even more domains in our future work,.

\subsection{Benchmark Characteristics}
\DomainCodeBench is designed with three key characteristics to allow comprehensive evaluation of LLMs in real-world development scenarios: multi-domain coverage, manually-annotated data, and dependency-enriched context.

\textbf{Multi-domain.} \DomainCodeBench spans 12 application domains, aiming to evaluate LLMs' code generation performance in real-world development scenarios. 
Previous benchmarks that focus on general-domain code generation tasks struggle to provide developers with the most direct guidance on which LLM is the strongest in their specific application domain. In contrast, \DomainCodeBench offers practical advice for developers, helping them choose the most suitable LLM in their respective domains. We advocate for a shift in LLMs' evaluation towards real-world scenarios, providing more direct feedback to researchers and practitioners.

\textbf{Manually-annotated.} 
Each task in \DomainCodeBench is accompanied by a manually written, high-quality docstring. 
During the benchmark construction process, we observe that the quality of docstrings in real-world projects varies significantly, with many failing to clearly and accurately convey the target function's purpose. Additionally, the human annotation process helps mitigate the risk of data leakage by providing fresh annotations, offering a more accurate reflection of the LLMs' code generation abilities in specific domains. Therefore, we deem it necessary to rewrite a high-quality docstring for each programming task.

\textbf{Dependency-enriched.} Each task in \DomainCodeBench is accompanied by rich relevant dependency information, which facilitates deeper analysis of LLMs' performance. We extract various dependencies related to the target function, including import statements from the local file and API dependencies of the target function. The APIs that the target function relies on are categorized into three types: standard library APIs, project-defined APIs, and third-party library APIs. This categorization allows for a more detailed analysis of LLMs, such as investigating the reasons behind LLMs' failures, assessing the LLMs' familiarity with these three types of APIs, and evaluating how well the LLMs handle domain-specific third-party libraries. This deeper insight aids in understanding the LLMs' strengths and weaknesses in handling complex, real-world tasks.

\begin{table}[ht]
\centering
\small
\setlength{\tabcolsep}{8pt} 
\caption{Performance comparison of LLMs on HumanEval.}
\label{table:Detailed-info-about-the-studied-LLMs}
\tabmargin
\begin{tabular}{llcc}
\toprule
\textbf{Type} & \textbf{Model}  & \textbf{Pass@1} &\textbf{Rank} \\
\midrule
Closed-source
& GPT-4  & 83.5 &1\\
\midrule
\multirow{9}{*}{Open-source} 
& CodeLLaMa-7B  & 39.6 &7 \\
& CodeLLaMa-13B  & 42.7 &6 \\
& CodeLLaMa-34B  & 52.4 &2 \\
& DeepSeekCoder-6.7B  & 45.1 &4 \\
 & DeepSeekCoder-33B  & 50.6 &3 \\
& StarCoder  & 34.8 &8 \\
& StarCoder2-3B  & 31.1  &10 \\
& StarCoder2-7B  & 34.8  &8 \\
& StarCoder2-15B   & 45.1 &4 \\

\bottomrule
\end{tabular}
\end{table}

\section{Experimental Setup}
\subsection{Research Questions}
We aim to answer the following research questions (RQs):
\begin{itemize}[leftmargin=10pt]
    \item\textbf{RQ1: Do LLMs with top performance in general domains necessarily achieve top performance across specific domains?} 
       
       To answer this question, we evaluate the code generation performance of ten mainstream LLMs on \DomainCodeBench to compare whether their performance rankings across different domains align with their rankings in general domains. 
    
    \item 
    \textbf{RQ2: What are the reasons that LLMs fail to correctly generate domain-specific code? }
       
       We conduct a manual error analysis on the generated results of LLMs and summarize the reasons for their failures in domain-specific code generation tasks.
    

    \item\textbf{RQ3: Can existing code generation methods effectively improve the domain-specific code generation performance of LLMs?}
      
      First, we classify existing repository-level code generation methods and then explore whether representative repository-level code generation methods can effectively improve the code generation performance of LLMs in specific domains.

\end{itemize}

\subsection{Experimental Settings}

\ \ \ \ \textbf{Model Selection.} We select the mainstream LLMs commonly used in recent code-related studies, including both open-source and closed-source LLMs. For open-source LLMs, we choose the StarCoder series~\cite{starcoder} (including StarCoder, StarCoder2-3B, StarCoder2-7B, and StarCoder2-15B), the CodeLLaMa series~\cite{codellama} (including CodeLLaMa-7B, CodeLLaMa-13B, and CodeLLaMa-34B), and the DeepSeekCoder series~\cite{deepseek} (including DeepSeekCoder-6.7B and DeepSeekCoder-33B). For closed-source LLM, we select the widely-used GPT-4~\cite{gpt} (GPT-4-0125-preview). Table~\ref{table:Detailed-info-about-the-studied-LLMs} provides detailed information about the studied LLMs.  
Note that LLMs' pass@1 scores on HumanEval are sourced from EvoEval~\cite{evoeval}.

\textbf{Model Settings.} Following prior work~\cite{evocodebench}, we set the temperature to 0.4 and top-p to 0.95. To mitigate randomness of model generation, the experimental results presented in this paper are obtained by conducting three repeated experiments and averaging the results. Besides, to better showcase LLMs' code generation performance, we combine the docstring and function signature in accordance with the characteristics of the task's corresponding programming language.
All experiments are conducted on a server running Ubuntu 18.04.6 LTS with 128 Intel(R) Xeon(R) Platinum 8336C @ 2.30GHz CPUs and 8 NVIDIA A800 80GB PCIe GPUs.

\begin{table*}[t]
\centering
\setlength{\tabcolsep}{5pt} 
\caption{Performance of LLMs on \DomainCodeBench. Abbreviations: \textbf{BC}: Blockchain, \textbf{CS}: Cloud Service, \textbf{DA}: Data Analysis, \textbf{DL}: Deep Learning, \textbf{DApp}: Desktop Application, \textbf{DSys}: Distributed System, \textbf{EApp}: Enterprise Application, \textbf{Game}: Game Development, \textbf{IoT}: Internet of Things, \textbf{Mob}: Mobile Application, \textbf{Rob}: Robotics, \textbf{Web}: Web Development.}
\label{table:LLMs-performance-on-DomainCodeBench}
\tabmargin
\resizebox{0.9\textwidth}{!}{%
\begin{tabular}{l*{12}{>{\centering\arraybackslash}p{0.75cm}} |>{\centering\arraybackslash}p{0.75cm} } 
\toprule
\textbf{LLMs} & \textbf{BC} & \textbf{CS} & \textbf{DA} & \textbf{DL} & \textbf{DApp} & \textbf{DSys} & \textbf{EApp} & \textbf{Game} & \textbf{IoT} & \textbf{Mob} & \textbf{Rob} & \textbf{Web} &\textbf{$\sigma^2$}\\
\midrule
GPT-4 & 46.83 & 39.10 & 39.02 & 36.47 & 43.15 & 40.58 & 36.40 & \cellcolor{blue!20}\textbf{41.13} & 39.46 & \cellcolor{blue!20}\textbf{46.33} & 42.26 & 34.45  &14.49\\
CodeLLaMa-7B & 43.24 & 39.28 & 33.34 & 33.34 & 40.52 & 38.47 & 35.76 & 35.50 & 38.20 & 42.78 & 40.32 & 34.58  &11.85\\
CodeLLaMa-13B & 47.91 & 41.47 & 36.04 & 35.69 & 42.80 & 40.11 & 37.05 & 35.66 & 40.64 & 44.39 & 42.84 & 35.33  &16.72\\
CodeLLaMa-34B & 48.90 & 42.57 & 37.80 & 38.25 & 43.31 & 42.14 & 37.80 & 35.38 & 41.09 & 46.07 & 43.56 & 36.80  &16.44\\
DeepSeekCoder-6.7B & 49.04 & 41.60 & 36.03 & 35.66 & 43.79 & 40.51 & 37.22 & 34.13 & 40.69 & 42.03 & 44.13 & 34.93  &\textbf{20.28}\\
DeepSeekCoder-33B & \cellcolor{blue!20}\textbf{52.75} & \cellcolor{blue!20}\textbf{44.32} & \cellcolor{blue!20}\textbf{39.65} & \cellcolor{blue!20}\textbf{39.89} & \cellcolor{blue!20}\textbf{46.29} & \cellcolor{blue!20}\textbf{42.27} & \cellcolor{blue!20}\textbf{39.64}& 39.71 & \cellcolor{blue!20}\textbf{43.45} & 44.33 & \cellcolor{blue!20}\textbf{48.54} & \cellcolor{blue!20}\textbf{38.88}  &18.15\\
StarCoder-15.5B & 40.17 & 39.02 & 34.10 & 37.19 & 37.37 & 35.32 & 32.84 & 35.91 & 37.17 & 30.45 & 35.97 & 30.41  &11.19\\
StarCoder2-3B & 40.21 & 37.60 & 31.97 & 33.92 & 35.16 & 32.38 & 30.87 & 34.01 & 35.31 & 31.91 & 36.74 & 29.09  &10.13\\
StarCoder2-7B & 41.20 & 39.57 & 34.10 & 35.22 & 38.03 & 33.49 & 33.32 & 36.43 & 37.10 & 34.70 & 37.81 & 30.60  &9.61\\
StarCoder2-15B & 41.67 & 41.92 & 37.06 & 37.67 & 39.33 & 36.71 & 34.87 & 36.00 & 39.79 & 34.53 & 40.82 & 33.83  &7.94\\
\bottomrule
\end{tabular}
}
\end{table*}

\subsection{Evaluation Metric}

We use CodeBLEU~\cite{codebleu} as the primary metric for evaluating code generation ability, which has been commonly used in prior work~\cite{domcoder, ComplexCodeEval}. CodeBLEU evaluates both textual and semantic match, making it suitable for DomainCodeBench's 12 domains and 15 PLs. Note that we opt out for execution-based metrics like Pass@k{} for several reasons including complexity, scale, and practicality.
First, tasks in \DomainCodeBench involve framework-specific logis (e.g., Solidity) and third-party libraries, requiring complex environments. 
Second, executing 2,400 tasks across diverse domains would incur extreme computational costs.
Finally, CodeBLEU enables efficient evaluation while capturing critical code measurement dimensions.

\begin{figure}[t]
    \centering
    \includegraphics[width=0.45\textwidth]{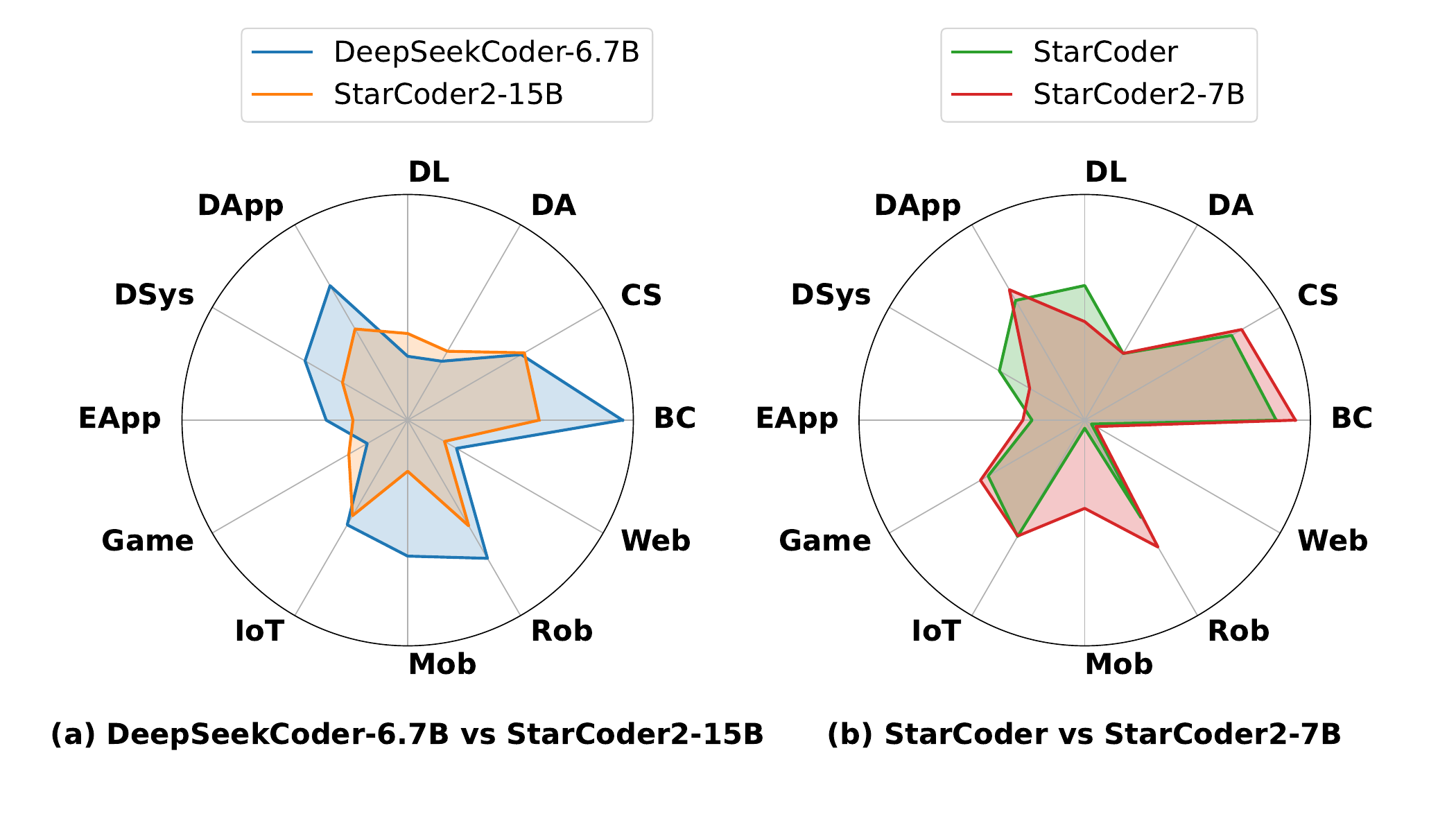} 
    \figmargin
    \figmargin
    \caption{Comparison of code generation performance in different domains for LLMs with similar performance on HumanEval.}
    \label{fig:code-gen-performance-Comparison} 
\end{figure}

\begin{table*}[t]
\centering
\small
\setlength{\tabcolsep}{7pt}
\caption{Code generation performance of LLMs in different sub-domains of enterprise application domain and game development domain. CL means CodeLLaMa; DSC means DeepSeekCoder; SC means StarCoder.}
\label{table:subdomain-performance}
\tabmargin
\begin{tabular}{lcclrccccccc}
\toprule
\textbf{Subdomain} & CL-7B & CL-13B & CL-34B & DSC-6.7B & DSC-33B & SC & SC2-3B & SC2-7B & SC2-1.5B & Mean & $\sigma^2$ \\
 \toprule
\rowcolor{blue!20} \multicolumn{12}{c}{Enterprise Application} \\
          ERP & 32.95 & 33.49 & 34.66 & 35.07 & \underline{36.75} & 31.79 & 29.53 & 31.52 & 33.31 &    33.23 &      4.07 \\
          CRM & 37.57 & 39.45 & 40.22 & 40.19 & \underline{41.95} & 31.15 & 31.15 & 33.76 & 34.07 &    36.61 &     \textbf{15.28} \\
          CMS & 32.17 & 36.59 & 34.80 & 35.51 & \underline{38.51} & 32.48 & 28.97 & 31.58 & 33.21 &    33.76 &      7.43 \\
          SCM & 41.00 & 42.36 & 43.24 & 40.87 & \underline{43.41} & 36.65 & 33.67 & 37.06 & 39.64 &    \textbf{39.77} &      9.89 \\
         HRMS & 31.91 & 30.64 & 31.79 & 29.20 & \underline{35.72} & 31.86 & 30.07 & 30.99 & 32.52 &    31.63 &      3.05 \\
         \hdashline 
 $\sigma^2$ & 12.86 & 17.29 & 17.35 & \textbf{17.7} & 8.77 & 3.91 & 2.76 & 5.06 & 6.72 & - & -\\
\hline
\rowcolor{blue!20} \multicolumn{12}{c}{Game Development} \\
     Unity3D & 36.58 & 38.40 & 40.56 & 38.68 & \underline{42.24} & 34.37 & 28.82 & 34.85 & 31.19 &    36.19 &     16.82 \\
 Unreal\_Engine & 34.13 & 29.90 & 26.80 & 24.89 & 35.24 & 42.36 & 40.38 & 40.84 & \underline{42.46} &    35.22 &     \textbf{40.97} \\
  Godot\_Engine & 38.90 & 36.64 & \underline{42.97} & 36.43 & 41.14 & 33.89 & 34.52 & 36.16 & 40.02 &    37.85 &      8.44 \\
    cocos2d-x & 38.72 & 39.75 & 40.65 & 42.63 & \underline{44.43} & 36.02 & 38.56 & 38.74 & 38.31 &    \textbf{39.76} &      5.61 \\
       libgdx & 31.92 & 34.87 & 32.80 & 32.62 & \underline{36.60} & 29.08 & 26.44 & 28.53 & 28.34 &    31.24 &     10.02 \\
       phaser & 34.11 & 39.60 & 38.01 & 36.58 & \underline{41.89} & 29.25 & 32.50 & 33.48 & 35.63 &    35.67 &     13.21 \\
\hdashline 
 $\sigma^2$ & 6.57 & 11.67 & \textbf{30.72} & 30.6 & 10.56 & 20.12 & 24.45 & 15.42 & 24.2 & - & -\\
\bottomrule
\end{tabular}

\end{table*}

\section{Evaluation Results}   
\subsection{Performance Decoupling (RQ1)}
Table~\ref{table:LLMs-performance-on-DomainCodeBench} presents the code generation performance of ten mainstream LLMs on \DomainCodeBench. The experimental results reveal a performance decoupling phenomenon: LLMs excelling in general domain do not consistently maintain superiority in specialized application domains. As shown in Table~\ref{table:Detailed-info-about-the-studied-LLMs}, DeepSeekCoder-33B ranks 4th on 
HumanEval yet achieves top performance in most application domains. 
In contrast, GPT-4 leads in general-domain performance but underperforms in domain-specific tasks.  
Similarly, while CodeLLaMA-34B surpasses DeepSeekCoder-33B on HumanEval, it trails in all application domains except  Mobile Application domain. 

Notably, all the evaluated LLMs exhibit substaintial performance gaps in \DomainCodeBench compared to general-domain benchmarks. Among these models, only DeepSeekCoder-33B exceeds 50 CodeBLEU score (blockchain domain), with even SOTA closed-source LLMs like GPT-4 demonstrating moderate domain-specific performance ($\leq 46.33$ CodeBLEU). These results underscore the distinct challenges posed by application-domain coding tasks compared to general domains. 


\begin{center}
    \begin{myboxc}{\textbf{Finding 1: }
    Superior performance on general-domain benchmarks does not guarantee equivalent capability in domain-specific code generation. Current LLMs exhibit substantial performance gaps in application-domain code generation.
    }
    \end{myboxc}
\end{center}

Moreover, by analyzing the performance variance of LLMs across different domains, we find that LLMs exhibit significant differences in performance across various domains. LLMs generally perform well in blockchain application development but show weaker performance in domains such as web development and enterprise application development. For example, DeepSeekCoder-6.7B achieves a CodeBLEU score of 49.04 in the Blockchain domain but only 34.93 in the Web development domain. 

\begin{center}
    \begin{myboxc}{\textbf{Finding 2:}
   LLMs demonstrate significant variations in code generation performance across different application domains, and the ten mainstream LLMs exhibit similar trends in their strengths and weaknesses across domains.
    }
    \end{myboxc}
\end{center}



As illustrated in Figure~\ref{fig:code-gen-performance-Comparison}, LLMs with similar performance on HumanEval can exhibit varying levels of performance in different application domains. For example, DeepSeekCoder-6.7B and StarCoder2-15B perform similarly in terms of code generation capability on HumanEval, yet their performance diverges significantly in different downstream application domains. DeepSeekCoder-6.7B excels in desktop application development, blockchain development, and mobile application development, while StarCoder2-15B outperforms DeepSeekCoder-6.7B in game development and deep learning. 
Similarly, as depicted in Figure~\ref{fig:code-gen-performance-Comparison} (b), StarCoder and StarCoder2-7B also show similar code generation performance on HumanEval. However, the domains in which these two LLMs excel are also different. 
Additionally, from the results in Table~\ref{table:LLMs-performance-on-DomainCodeBench}, we observe that the larger parameter scale of LLMs does not necessarily correlate with better code generation performance. For example, the 15.5B StarCoder performs significantly worse than the 7B CodeLLaMa model in several domains, including blockchain development, cloud services, desktop application development, enterprise application development, IoT, mobile application development, robotics, and web development.
\begin{center}
    \begin{myboxc}{\textbf{Finding 3:}
   LLMs that exhibit similar performance in general domains show significant performance differences in various application domains. Besides, a larger model does not necessarily achieve better code generation performance across all application domains.
    }
    \end{myboxc}
\end{center}

   

\textbf{Performance Comparison Among Subdomains.} We further investigate the performance of LLMs across different sub-domains to better explain their performance in various practical development scenarios. Due to the space limitation of the paper, Table~\ref{table:subdomain-performance} presents the performance of LLMs in the Enterprise Application development domain and the Game development domain. The complete experimental data are available in the online appendix within our repository~\cite{DomainCodeBench}. Overall, the performance of LLMs across different sub-domains within the same domain varies significantly. The average CodeBLEU score for all models in HRMS system development is 31.63, while the score in SCM system development is 39.77, which is 25.74\% higher. Additionally, different LLMs show the greatest variance in CodeBLEU scores in the CRM system development sub-domain within Enterprise Applications, indicating that there are considerable differences in performance even within the same sub-domain. Furthermore, from the performance of individual models, we can also see significant variations in their performance across different sub-domains. For example, DeepSeekcoder-6.7B shows substantial fluctuations in code generation performance across different sub-domains within the Enterprise Application domain, while StarCoder demonstrates more stable performance.

\begin{center}
    \begin{myboxc}{\textbf{Finding 4:}
  LLMs exhibit significant performance differences across different sub-domains within the same domain. And within the same sub-domain, different LLMs show considerable performance variations.
    }
    \end{myboxc}
\end{center}

\begin{table*}[h]
    \centering
    \footnotesize
    \renewcommand{\arraystretch}{1} 
    \caption{Failure taxonomy of domain-specific code generation. FOC stands for frequency of occurrence.}
    \label{table:reason_category}
    \tabmargin
    \tabmargin
    \begin{tabular}{|>{\centering\arraybackslash}m{1.7cm}| 
                    >{\centering\arraybackslash}m{4.7cm} | 
                    >{\centering\arraybackslash}m{8.5cm} | 
                    >{\centering\arraybackslash}m{0.8cm}|}
        \hline
        \textbf{Category} & \textbf{Sub-category} & \textbf{Description} & \textbf{FOC} \\
        \hline
        \multirow{3}{*}{\parbox{1.7cm}{\centering Domain-specific \\ Reason}}
        & Insufficient understanding of project background 
        & LLM generates code that cannot properly call custom APIs or fails to meet project specifications due to lack of understanding of the specific domain project's background. 
        & 0.97 \\ \cline{2-4}
        & Limited understanding of domain-specific libraries 
        & LLM struggles to generate correct code because it is unfamiliar with or unable to correctly call the APIs of third-party libraries specific to the domain. 
        & 0.77 \\ \cline{2-4}
        & Unfamiliar with domain-specific algorithms or workflows 
        & LLM generates code that fails to meet domain-specific requirements due to unfamiliarity with commonly used algorithms or workflows in the domain. 
        & 0.66 \\ 
        \hline
        \multirow{2}{*}{\centering General Reason} 
        & Insufficient requirement comprehension & LLM generates code that does not correctly implement the intended functionality due to insufficient comprehension of the coding requirements. 
        & 0.50 \\ \cline{2-4}
        & Limited proficiency in low-resource programming languages 
        & LLM's coding ability in less commonly used programming languages is weak due to a lack of training data for these languages, often resulting in incomplete or repetitive code statements. 
        & 0.02 \\ 
        \hline
    \end{tabular}
    
\end{table*}


Based on the experimental results and analysis above, we call on researchers to shift their focus from general-domain benchmarks to more specific downstream domains, which will provide deeper insights into LLM performance and offer more practical, actionable guidance for users of these LLMs. Instead of simply reporting a single pass@k score~\cite{humaneval}, \DomainCodeBench allows detailed analysis for LLMs' code generation capabilities across different software application development domains to identify specific domain where LLMs struggle or excel.

\subsection{Root Causes of Failures (RQ2)}
\label{fail-analysis}

\ \ \ \ \ \textbf{Taxonomy of Failure Casues.} Given that all the studied LLMs show unsatisfactory performance in domain-specific code generation tasks, we conduct a manual analysis of their generated outputs to investigate the underlying causes of errors and assess the severity of different causes (i.e., the frequency of their occurrence). This error analysis process involves five annotators, each with more than five years of programming experience. Specifically, we randomly select 10\% of the generated outputs from all the LLMs and require three annotators to individually analyze the reasons for errors in each sample. It is important to note that a single sample may have multiple error causes. Afterward, we summarize and categorize the causes of these errors. Then, all annotators analyze the generated outputs from all the LLMs based on this classification. If the cause of an error in a generated output matches one of the existing categories, we increase the count for that category. If the error cause does not belong to any existing category, we add a new category and re-annotate all the outputs accordingly.

As shown in Table~\ref{table:reason_category}, after extensive manual error analysis, we identify five key reasons that contribute to the LLM's inability to generate the domain-specific code. We categorize the reasons into two major categories: domain-specific reasons and general reasons related to the model's general capabilities. Among these, domain-specific reasons are the most significant. Therefore, in the following content, we will primarily focus on analyzing domain-specific factors.

\begin{center}
    \begin{myboxc}{\textbf{Finding 5:}
  We conduct an in-depth investigation into the failure casuses of LLMs in domain-specific code generation tasks and construct a taxonomy of failure causes, offering guidance for further improving LLMs' capabilities.
    }
    \end{myboxc}
\end{center}

\begin{figure*}[ht] 
    \centering
    \includegraphics[width=1\textwidth]{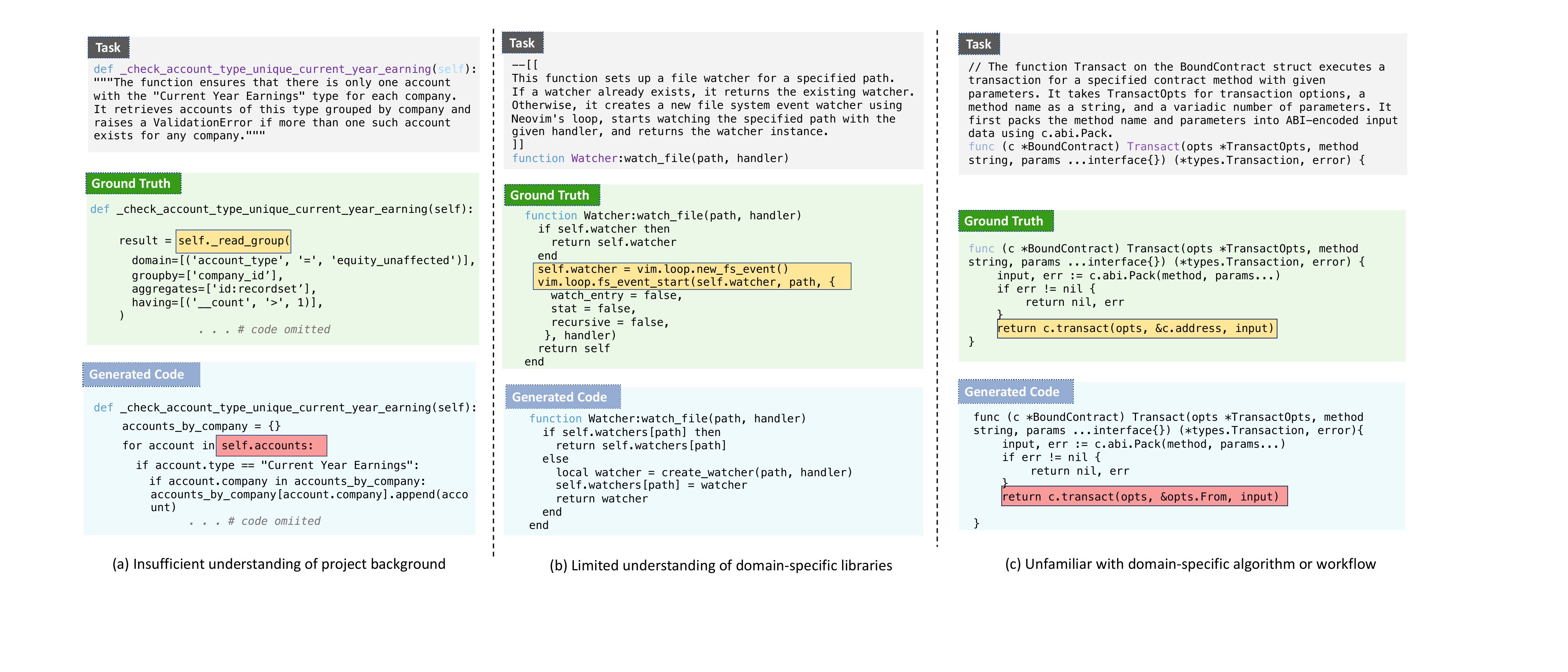} 
    \figmargin
    \figmargin
    \figmargin
    \figmargin
    \caption{Case studies.}
    \label{fig:case-study} 
    
\end{figure*}

\textbf{Case Study 1: Insufficient understanding of project background.} The most significant reason behind the LLMs' inability to generate target code correctly is their insufficient understanding of the project background. Real-world projects are often intricate and have complex dependencies, so developers must understand the interdependencies between programs to write code. This also applies to LLMs, as many functions defined in the project may not have been encountered during their training, leading to a lack of awareness of these dependency functions. Consequently, LLMs may perform poorly in real project development scenarios due to a lack of understanding of the relevant project context. Figure~\ref{fig:case-study} (a) presents a case where GPT-4 misuses an API due to a lack of repository context. The implementation of the ground truth requires querying a database defined within the project, which has corresponding database access interfaces ``self.read\_group''. However, since the LLM is unaware of these defined interfaces, it fails to complete the code generation task correctly and resorts to accessing an undefined ``self.accounts'' interface.
\begin{center}
    \begin{myboxc}{\textbf{Finding 6:}
  During generating domain-specific code, LLMs often struggle to generate correct code due to insufficient understanding of project-specific context, such as dependencies and defined functions, leading to errors like misusing APIs or accessing undefined interfaces.
    }
    \end{myboxc}
\end{center}

\textbf{Case Study 2: Limited understanding of domain-specific libraries.} It is well known that different development domains may have various third-party libraries that accelerate software development by providing APIs. However, due to the large number of these libraries and the multiple parameter constraints associated with different APIs, even if the LLM has encountered data related to these APIs during training, it can still struggle to use them flexibly~\cite{domaineval,domcoder}. Figure~\ref{fig:case-study} (b) presents a case in which the LLM fails to call third-party APIs. The implementation of the ground truth requires calling two APIs provided by a third-party library to create a new file system event watcher and utilize it to monitor file system events at a specific path. However, GPT-4 fails to invoke these APIs, leading to errors. 
\begin{center}
    \begin{myboxc}{\textbf{Finding 7:}
  Due to the vast number of third-party libraries and the complex parameter constraints of different APIs, LLMs struggle to effectively utilize them, often failing to correctly invoke these APIs even if LLMs have encountered them during training.
    }
    \end{myboxc}
\end{center}

\textbf{Case Study 3: Unfamiliarity with domain-specific algorithms or workflow}. As shown in Figure~\ref{fig:case-study} (c), LLM fails to correctly distinguish between the contract address and the transaction sender's address when executing contract transactions due to its lack of familiarity with blockchain domain knowledge, resulting in incorrect results. To achieve proficiency in domain-specific algorithms and processes, LLMs should be trained on a large amount of domain-specific data, which allows them to effectively learn and derive domain-specific processing workflows from the data. In real development, it is unrealistic to expect programmers to provide detailed explanations of foundational knowledge in their development domain whenever they use the LLM to generate code, as this would waste a significant amount of their time and effort.

\begin{center}
    \begin{myboxc}{\textbf{Finding 8:}
  LLMs frequently fail in code generation tasks due to insufficient familiarity with domain-specific algorithms and processing methods, highlighting the necessity of specialized domain training data.
    }
    \end{myboxc}
\end{center}

\begin{table*}[t]
\centering
\footnotesize
\setlength{\tabcolsep}{3pt} 
\caption{Performance of LLMs on various domains with different contexts. }
\tabmargin
\tabmargin
\label{table:llms_performance_multiple_context}
\begin{tabular}{l*{12}{>{\centering\arraybackslash}p{0.9cm}}  >{\centering\arraybackslash}p{0.9cm}} 
\toprule
\textbf{LLMs} & \textbf{BC} & \textbf{CS} & \textbf{DA} & \textbf{DL} & \textbf{DApp} & \textbf{DSys} & \textbf{EApp} & \textbf{Game} & \textbf{IoT} & \textbf{Mob} & \textbf{Rob} & \textbf{Web} &\textbf{Mean}\\

\toprule
\rowcolor{blue!20} \multicolumn{14}{c}{API Context} \\

CodeLLaMa-7B & 5.90 & 2.45 & 1.20 & 3.54 & 5.95 & 1.75 & 2.89 & 7.23 & 3.58 & 0.05 & 4.42 & 3.84 & 3.57 \\
CodeLLaMa-13B & 5.45 & 3.21 & 1.27 & 3.34 & 7.49 & 2.96 & 3.15 & 6.83 & 4.55 & 1.16 & 5.36 & 4.70 & 4.12 \\
CodeLLaMa-34B & 4.89 & 2.47 & 0.71 & 2.64 & 6.68 & 1.71 & 2.36 & 8.59 & 4.04 & 0.44 & 5.70 & 3.96 & 3.68 \\
DeepSeekCoder-6.7B & 5.66 & 1.69 & 1.95 & 2.02 & 5.57 & 2.13 & 2.74 & 9.27 & 2.84 & -0.78 & 3.94 & 4.65 & 3.47 \\
DeepSeekCoder-33B & 4.94 & 2.27 & 1.67 & 4.48 & 6.94 & 3.16 & 3.53 & 8.55 & 3.81 & 0.35 & 5.11 & 4.38 & 4.10 \\
StarCoder & 5.96 & 2.40 & 1.58 & 2.67 & 3.50 & 1.52 & 2.79 & 4.50 & 2.91 & 6.64 & 4.25 & 4.29 & 3.58 \\
StarCoder2-3B & 4.06 & -0.11 & 1.96 & 0.45 & 5.17 & 0.06 & 2.15 & 5.06 & 2.16 & 2.16 & 3.87 & 3.08 & 2.51 \\
StarCoder2-7B & 3.01 & 0.10 & 0.22 & 0.06 & 3.57 & 0.51 & 0.50 & 3.97 & 3.00 & 0.69 & 3.75 & 2.87 & 1.85 \\
StarCoder2-15B & 7.55 & 0.96 & 2.21 & 2.43 & 5.75 & 2.99 & 3.08 & 7.53 & 3.37 & 5.96 & 4.44 & 3.95 & 4.18 \\

\midrule
\rowcolor{blue!20} \multicolumn{14}{c}{Dependency Context} \\

CodeLLaMa-7B & 6.34 & 4.12 & 5.22 & 3.52 & 7.98 & 6.75 & 6.02 & 7.32 & 5.31 & 11.76 & 2.01 & 6.56 & 6.08 \\
CodeLLaMa-13B & 4.81 & 3.60 & 4.08 & 4.59 & 8.31 & 7.13 & 6.72 & 4.22 & 3.74 & 11.22 & 2.01 & 7.14 & 5.63 \\
CodeLLaMa-34B & 4.29 & 2.59 & 2.41 & 2.98 & 7.56 & 7.73 & 6.35 & 4.89 & 3.67 & 10.81 & 2.17 & 6.94 & 5.20 \\
DeepSeekCoder-6.7B & 4.01 & 3.48 & 3.96 & 3.29 & 7.11 & 7.01 & 7.28 & 3.49 & 2.91 & 10.33 & -0.78 & 7.57 & 4.97 \\
DeepSeekCoder-33B & 4.39 & 3.37 & 4.58 & 4.37 & 7.29 & 7.57 & 7.39 & 1.93 & 3.01 & 10.65 & 0.12 & 6.61 & 5.11 \\
StarCoder & 7.13 & 4.80 & 3.13 & 3.35 & 5.68 & 6.60 & 6.37 & 1.21 & 1.73 & 10.37 & 1.65 & 5.57 & 4.80 \\
StarCoder2-3B & 1.96 & 2.70 & 5.67 & 2.34 & 5.71 & 5.75 & 6.11 & 2.31 & 2.04 & 10.03 & -1.08 & 3.34 & 3.91 \\
StarCoder2-7B & 4.18 & 2.35 & 5.74 & 2.99 & 5.28 & 5.44 & 5.93 & 2.85 & 2.60 & 6.16 & -1.04 & 3.91 & 3.87 \\
StarCoder2-15B & 6.11 & 2.58 & 5.58 & 3.67 & 3.92 & 6.54 & 6.75 & 0.57 & 1.48 & 10.50 & -0.59 & 4.80 & 4.33 \\

\midrule
\rowcolor{blue!20} \multicolumn{14}{c}{Similar Context} \\

CodeLLaMa-7B & 13.95 & 12.23 & 4.12 & 5.78 & 10.82 & 11.50 & 8.19 & 14.35 & 10.31 & 11.56 & 14.79 & 8.80 & 10.52 \\
CodeLLaMa-13B & 11.90 & 9.45 & 4.43 & 6.41 & 11.80 & 13.36 & 8.27 & 14.43 & 11.31 & 13.12 & 14.20 & 10.56 & 10.63 \\
CodeLLaMa-34B & 11.19 & 10.76 & 3.62 & 5.87 & 10.48 & 10.57 & 7.81 & 16.91 & 9.84 & 12.11 & 13.59 & 10.72 & 10.47 \\
DeepSeekCoder-6.7B & 13.30 & 12.93 & 6.16 & 6.52 & 11.48 & 14.41 & 9.61 & 21.63 & 12.18 & 15.34 & 15.06 & 11.72 & 12.38 \\
DeepSeekCoder-33B & 12.07 & 12.75 & 6.82 & 7.46 & 12.17 & 14.21 & 9.91 & 18.46 & 11.83 & 15.16 & 12.08 & 10.88 & 11.71 \\
StarCoder & 15.51 & 11.85 & 5.34 & 6.50 & 13.15 & 13.98 & 10.18 & 13.53 & 11.97 & 20.58 & 17.52 & 10.20 & \textbf{12.65} \\
StarCoder2-3B & 9.21 & 8.97 & 6.33 & 6.49 & 6.53 & 11.47 & 7.66 & 13.18 & 9.91 & 9.96 & 12.74 & 2.87 & 7.92 \\
StarCoder2-7B & 10.68 & 8.62 & 6.82 & 5.95 & 8.75 & 10.82 & 6.90 & 11.68 & 9.69 & 14.06 & 13.08 & 5.74 & 8.71 \\
StarCoder2-15B & 8.10 & 9.67 & 5.58 & 7.17 & 4.06 & 10.83 & 7.25 & 6.39 & 7.22 & 13.08 & 9.40 & 3.35 &7.24 \\

\midrule
\rowcolor{blue!20} \multicolumn{14}{c}{API + Dependency Context} \\
CodeLLaMa-7B & 9.82 & 4.87 & 5.73 & 5.33 & 12.12 & 7.80 & 8.99 & 9.96 & 6.15 & 13.06 & 6.14 & 7.68 & 7.96 \\
CodeLLaMa-13B & 6.77 & 5.15 & 5.64 & 5.45 & 13.49 & 9.43 & 8.99 & 10.69 & 6.63 & 11.58 & 6.38 & 8.37 & 7.76 \\
CodeLLaMa-34B & 7.48 & 5.53 & 5.48 & 5.28 & 12.80 & 8.58 & 8.70 & 11.83 & 6.82 & 11.63 & 7.23 & 8.73 & 8.07 \\
DeepSeekCoder-6.7B & 6.86 & 5.05 & 6.67 & 6.97 & 11.73 & 7.74 & 9.14 & 11.95 & 5.14 & 10.87 & 4.41 & 7.87 & 7.33 \\
DeepSeekCoder-33B & 6.47 & 5.69 & 5.33 & 7.11 & 12.37 & 9.09 & 10.27 & 11.86 & 6.13 & 11.39 & 4.70 & 7.74 & 7.39 \\
StarCoder & 7.10 & 5.18 & 6.43 & 6.16 & 5.75 & 6.56 & 6.89 & 4.06 & 2.68 & 7.50 & 3.22 & 4.74 & 5.57 \\
StarCoder2-3B & 5.23 & 3.78 & 7.10 & 4.34 & 8.36 & 5.71 & 8.53 & 7.94 & 3.62 & 8.63 & 2.73 & 3.72 & 5.53 \\
StarCoder2-7B & 4.09 & 2.06 & 6.63 & 2.20 & 6.54 & 3.80 & 5.14 & 5.14 & 4.25 & 5.01 & 2.38 & 3.67 & 4.53 \\
StarCoder2-15B & 7.56 & 4.56 & 6.97 & 5.60 & 9.49 & 8.15 & 9.18 & 9.14 & 5.19 & 11.18 & 4.63 & 4.61 &7.19 \\

\midrule
\rowcolor{blue!20} \multicolumn{14}{c}{API + Similar Context} \\
CodeLLaMa-7B   & 15.68 & 14.28 & 6.05 & 7.97 & 15.73 & 12.82 & 12.34 & 19.05 & 13.94 & 13.94 & 17.85 & 11.81 & 13.94 \\
CodeLLaMa-13B  & 14.66 & 13.42 & 6.88 & 6.43 & 16.84 & 14.93 & 12.10 & 21.51 & 13.33 & 12.71 & 18.23 & 13.64 & 13.56 \\
CodeLLaMa-34B  & 13.22 & 13.62 & 4.92 & 7.49 & 16.83 & 11.98 & 11.84 & 23.03 & 12.89 & 11.91 & 17.05 & 13.36 & 13.07 \\
DeepSeekCoder-6.7B  & 14.82 & 14.66 & 7.15 & 7.67 & 15.63 & 15.07 & 12.00 & 24.30 & 13.34 & 14.07 & 17.75 & 13.76 & 14.63 \\
DeepSeekCoder-33B   & 14.06 & 14.07 & 7.01 & 8.36 & 15.86 & 15.85 & 12.53 & 21.37 & 14.65 & 15.10 & 17.21 & 13.32 & 14.23 \\
StarCoder     & 14.70 & 13.15 & 7.00 & 7.17 & 12.36 & 13.64 & 9.45 & 14.07 & 11.37 & 8.10 & 15.40 & 5.31 & 11.13 \\
StarCoder2-3B & 8.81 & 9.68 & 6.68 & 5.72 & 8.42 & 6.97 & 8.47 & 11.62 & 10.24 & 8.40 & 13.17 & -0.18 & 7.93 \\
StarCoder2-7B & 9.62 & 9.53 & 6.89 & 5.27 & 9.07 & 8.11 & 7.91 & 9.61 & 10.48 & 7.56 & 12.29 & 3.11 & 7.99 \\
StarCoder2-15B & 12.26 & 10.52 & 7.89 & 8.30 & 13.32 & 12.33 & 9.33 & 13.58 & 11.09 & 13.31 & 13.07 & 4.34 & 10.63 \\

\midrule
\rowcolor{blue!20} \multicolumn{14}{c}{API + Similar Context + Dependency Context} \\

CodeLLaMa-7B & 16.91 & 15.49 & 9.97 & 10.50 & 21.45 & 19.13 & 17.00 & 21.48 & 16.56 & 23.44 & 18.36 & 15.42 & \textbf{17.14} \\
CodeLLaMa-13B & 15.52 & 15.15 & 10.08 & 9.27 & 22.03 & 18.80 & 15.93 & 24.15 & 15.99 & 23.24 & 18.83 & 16.79 & \textbf{17.15} \\
CodeLLaMa-34B & 14.50 & 15.51 & 8.76 & 9.38 & 21.59 & 16.92 & 16.22 & 25.09 & 15.28 & 21.23 & 18.53 & 16.00 & \textbf{16.58} \\
DeepSeekCoder-6.7B & 15.71 & 17.72 & 10.56 & 10.32 & 20.40 & 18.89 & 16.55 & 27.55 & 15.48 & 23.02 & 18.48 & 17.00 & \textbf{17.64} \\
DeepSeekCoder-33B & 14.95 & 15.91 & 10.19 & 11.58 & 21.05 & 20.21 & 16.77 & 24.72 & 16.77 & 24.02 & 17.79 & 17.51 & \textbf{17.62} \\
StarCoder & 12.52 & 12.27 & 10.18 & 9.45 & 10.16 & 10.82 & 11.57 & 14.67 & 11.54 & 11.67 & 17.05 & 4.52 & 11.37 \\
StarCoder2-3B & 10.17 & 12.17 & 10.73 & 8.57 & 11.09 & 11.48 & 11.26 & 13.40 & 10.77 & 13.04 & 13.30 & 2.47 & \textbf{10.70} \\
StarCoder2-7B & 11.12 & 11.17 & 10.24 & 7.78 & 12.32 & 12.07 & 10.13 & 13.26 & 12.26 & 15.56 & 12.57 & 4.48 & \textbf{11.08} \\
StarCoder2-15B & 13.89 & 12.47 & 12.05 & 10.50 & 17.12 & 14.79 & 12.88 & 16.19 & 14.06 & 20.13 & 12.44 & 7.65 & \textbf{13.68} \\

\bottomrule
\end{tabular}
    \figmargin
\end{table*}

\subsection{Improvement Possibility (RQ3)}

We investigate whether existing code generation methods can effectively enhance the performance of LLMs in domain-specific tasks. Since developers typically write domain-specific code within a particular code repository, we focus on repository-level code generation methods~\cite{repocoder,repositorylevel,codegen4libs,ComplexCodeEval,ding2022cocomic,wang2024rlcoder}.

\textbf{Experimental Setup.} 
Current methods can be categorized into two main strategies: retrieval-based methods and static analysis-based methods. 
We cover both of them to comprehensively reveal the role of existing code generation methods in specific domains. 
Retrieval-based methods~\cite{repocoder} aim to extract code snippets from the repository that exhibit high similarity to the target code to aid LLMs in code generation. In this experiment, we adopt a standard retrieval-augmented generation (RAG) pipeline to retrieve code snippets from the repository that are likely highly similar to the target code (referred to as \texttt{Similar Context} in Table~\ref{table:llms_performance_multiple_context}). 
Static analysis-based methods typically involve parsing potential dependencies of the target function and extracting relevant code snippets from the repository. In this experiment, we directly extract the oracle project dependency context from the ground truth and provide these code snippets that the target function depends on (referred to as \texttt{Dependency Context} in Table~\ref{table:llms_performance_multiple_context}) to the LLM to explore the upper bound of the model's performance when given sufficient and accurate project dependency context information.

Since most existing repository-level code generation methods do not involve the mining of domain-specific information, 
we directly extract the third-party library API information required by the target function (referred to as \texttt{API Context} in Table~\ref{table:llms_performance_multiple_context}) from the ground truth and provide it as domain-specific input for the model. This allows us to explore LLMs' performance when given oracle domain-specific API information for the target domain.

\textbf{Analysis.} Table~\ref{table:llms_performance_multiple_context} shows a comparison of domain-specific code generation performance between the LLMs using the basic prompt and the LLMs using the basic prompt with additional information. The values in Table~\ref{table:llms_performance_multiple_context} are computed by subtracting the performance of the LLMs using the basic prompt from that of the LLMs using the basic prompt with additional information. Positive values indicate the extent to which the performance of LLMs improves when given additional information, while negative values indicate the extent to which the performance decreases under the same conditions. When we feed these three types of additional information (API context, dependency context, and similar context) to LLMs separately, the performance of domain-specific code generation improves to some extent in each case. Among these, the improvement with similar context is the most significant.

Next, we combine project context information with domain-specific knowledge (API context), finding that LLMs' domain-specific code generation performance improved across all models. From the experimental results, we find that when the input contains similar context, the performance improvement is relatively more pronounced. This is unexpected because the accuracy of similar context retrieved through search is generally limited, making it difficult to recall all the dependency information needed by the model. In the ``API + Dependency'' experiment, we provide the model with input that already contains all the dependencies and domain-specific third-party library knowledge required for the target code. However, LLMs perform better in the experiment with the only similar context than in the one where all dependencies and domain-specific third-party library knowledge are provided. Through data analysis, we discover that the model's ability to use domain-specific APIs is limited. In other words, even when told which APIs need to be called, the model still struggles to generate code snippets that correctly invoke these domain-specific APIs or follow domain-specific workflows. On the other hand, the similar context retrieves code snippets from the project that call the same APIs as the target code or follow similar workflows. These similar code snippets provide a good example for the LLMs, helping the model more directly generate code that meets domain-specific requirements. These LLMs, which perform well in general domains, struggle to effectively use domain-specific knowledge, which limits their performance in domain-specific code generation to some extent.

\begin{center}
    \begin{myboxc}{\textbf{Finding 9:}
    Providing LLMs with additional contexts can significantly enhance their code generation performance across various domains. Among these, similar context yields the most substantial performance improvement. Similar code snippets act as practical examples, allowing LLMs to perform better when referencing similar contexts.
    }
    \end{myboxc}
\end{center}

\section{Threats to validity}
\textbf{Internal Threats.} The first internal threat is the number of domains covered by \DomainCodeBench. \DomainCodeBench encompasses 12 specific application domains. However, due to the vast number of domains, it is challenging to cover all of them at once, which may somewhat limit the applicability of \DomainCodeBench. Additionally, our research on specific domains from active websites may overlook some equally prominent domains. Therefore, we continue to expand \DomainCodeBench in future work to promote its adoption across more domains. The second internal threat is the number of LLMs involved in our experiments. Due to computational resources limitations, we are unable to evaluate all code-related LLMs~\cite{li2022competition,nijkamp2022codegen,ahmad2021unified,chowdhery2023palm,fried2022incoder,zheng2023codegeex,wizardcoder,magicoder,pangu,codet,muennighoff2023octopack,qwen,chen2024identifying,chen2024rmcbench,zhang2024llm,zhong2024memorybank}. Nonetheless, we make every effort to cover both open-source and closed-source LLMs to enhance the representativeness of the experimental results. The third internal threat is the evaluation metric used in our experiments. We choose CodeBLEU as the evaluation metric for all experiments. Although this metric reflects the accuracy of LLM-generated results from various aspects, it does not fully capture whether the generated code can execute correctly or pass the corresponding test cases.

\textbf{External Threats.} The first external threat is that \DomainCodeBench contains data from GitHub before 2023, which may have already been included in the training datasets of the evaluated models. To mitigate this risk, we provide each task instance in \DomainCodeBench with a manually annotated docstring, avoiding the use of docstrings from online code repositories to prevent LLMs from directly associating training data with the programming tasks in \DomainCodeBench. The second external threat is the uncertainty in the model generation process. The inherent randomness in LLM generation may make it difficult to accurately reveal the models' code generation performance in specific domains. To mitigate this threat, we repeat each experiment three times and take the average result to reduce the impact of model uncertainty. The third external threat is prompt template selection. We ultimately choose the prompt template that yields the best results after testing several options. However, due to the vast number of available prompt templates, it is nearly impossible to cover all of them, which may slightly affect the representativeness of our experiments.

\section{Conclusion}

In this paper, we highlight the importance of evaluating LLMs in specific domains and introduce \DomainCodeBench, a multi-domain, human-verified, dependency-enriched code generation benchmark. \DomainCodeBench covers 12 prevalent application domains, each of which includes 200 meticulously curated programming tasks. We conduct comprehensive experiments to reveal the code generation performance of LLMs in different domains. Additionally, we delve into the underlying reasons behind LLMs' failures in code generation tasks, identifying key factors that contribute to their suboptimal performance in specific domains. We also investigate the effectiveness of existing code generation methods in assisting the accurate generation of domain-specific code. Experimental results provide valuable actionable insights, which we believe might assist developers in selecting more suitable code LLMs and help researchers further enhance these LLMs' coding abilities.

\bibliographystyle{ACM-Reference-Format}
\bibliography{ref}


\begin{thebibliography}{95}


\ifx \showCODEN    \undefined \def \showCODEN     #1{\unskip}     \fi
\ifx \showDOI      \undefined \def \showDOI       #1{#1}\fi
\ifx \showISBNx    \undefined \def \showISBNx     #1{\unskip}     \fi
\ifx \showISBNxiii \undefined \def \showISBNxiii  #1{\unskip}     \fi
\ifx \showISSN     \undefined \def \showISSN      #1{\unskip}     \fi
\ifx \showLCCN     \undefined \def \showLCCN      #1{\unskip}     \fi
\ifx \shownote     \undefined \def \shownote      #1{#1}          \fi
\ifx \showarticletitle \undefined \def \showarticletitle #1{#1}   \fi
\ifx \showURL      \undefined \def \showURL       {\relax}        \fi
\providecommand\bibfield[2]{#2}
\providecommand\bibinfo[2]{#2}
\providecommand\natexlab[1]{#1}
\providecommand\showeprint[2][]{arXiv:#2}

\bibitem[Achiam et~al\mbox{.}(2023)]%
        {gpt}
\bibfield{author}{\bibinfo{person}{Josh Achiam}, \bibinfo{person}{Steven Adler}, \bibinfo{person}{Sandhini Agarwal}, \bibinfo{person}{Lama Ahmad}, \bibinfo{person}{Ilge Akkaya}, \bibinfo{person}{Florencia~Leoni Aleman}, \bibinfo{person}{Diogo Almeida}, \bibinfo{person}{Janko Altenschmidt}, \bibinfo{person}{Sam Altman}, \bibinfo{person}{Shyamal Anadkat}, {et~al\mbox{.}}} \bibinfo{year}{2023}\natexlab{}.
\newblock \showarticletitle{Gpt-4 technical report}.
\newblock \bibinfo{journal}{\emph{arXiv preprint arXiv:2303.08774}} (\bibinfo{year}{2023}).
\newblock


\bibitem[Ahmad et~al\mbox{.}(2021)]%
        {ahmad2021unified}
\bibfield{author}{\bibinfo{person}{Wasi~Uddin Ahmad}, \bibinfo{person}{Saikat Chakraborty}, \bibinfo{person}{Baishakhi Ray}, {and} \bibinfo{person}{Kai-Wei Chang}.} \bibinfo{year}{2021}\natexlab{}.
\newblock \showarticletitle{Unified pre-training for program understanding and generation}.
\newblock \bibinfo{journal}{\emph{arXiv preprint arXiv:2103.06333}} (\bibinfo{year}{2021}).
\newblock


\bibitem[Alnaeli et~al\mbox{.}(2016)]%
        {alnaeli2016evolution}
\bibfield{author}{\bibinfo{person}{Saleh~M Alnaeli}, \bibinfo{person}{Melissa Sarnowski}, \bibinfo{person}{Md~Sayedul Aman}, \bibinfo{person}{Kumar Yelamarthi}, \bibinfo{person}{Ahmed Abdelgawad}, {and} \bibinfo{person}{Haowen Jiang}.} \bibinfo{year}{2016}\natexlab{}.
\newblock \showarticletitle{On the evolution of mobile computing software systems and C/C++ vulnerable code: Empirical investigation}. In \bibinfo{booktitle}{\emph{2016 IEEE 7th Annual Ubiquitous Computing, Electronics \& Mobile Communication Conference (UEMCON)}}. IEEE, \bibinfo{pages}{1--7}.
\newblock


\bibitem[Ars Technica({[n.\,d.]})]%
        {arstechnica}
Ars Technica \bibinfo{year}{[n.\,d.]}\natexlab{}.
\newblock
\newblock
\urldef\tempurl%
\url{https://arstechnica.com/information-technology/}
\showURL{%
\tempurl}


\bibitem[Athiwaratkun et~al\mbox{.}(2022)]%
        {athiwaratkun2022multi}
\bibfield{author}{\bibinfo{person}{Ben Athiwaratkun}, \bibinfo{person}{Sanjay~Krishna Gouda}, \bibinfo{person}{Zijian Wang}, \bibinfo{person}{Xiaopeng Li}, \bibinfo{person}{Yuchen Tian}, \bibinfo{person}{Ming Tan}, \bibinfo{person}{Wasi~Uddin Ahmad}, \bibinfo{person}{Shiqi Wang}, \bibinfo{person}{Qing Sun}, \bibinfo{person}{Mingyue Shang}, {et~al\mbox{.}}} \bibinfo{year}{2022}\natexlab{}.
\newblock \showarticletitle{Multi-lingual evaluation of code generation models}.
\newblock \bibinfo{journal}{\emph{arXiv preprint arXiv:2210.14868}} (\bibinfo{year}{2022}).
\newblock


\bibitem[Austin et~al\mbox{.}(2021)]%
        {mbpp}
\bibfield{author}{\bibinfo{person}{Jacob Austin}, \bibinfo{person}{Augustus Odena}, \bibinfo{person}{Maxwell Nye}, \bibinfo{person}{Maarten Bosma}, \bibinfo{person}{Henryk Michalewski}, \bibinfo{person}{David Dohan}, \bibinfo{person}{Ellen Jiang}, \bibinfo{person}{Carrie Cai}, \bibinfo{person}{Michael Terry}, \bibinfo{person}{Quoc Le}, {et~al\mbox{.}}} \bibinfo{year}{2021}\natexlab{}.
\newblock \showarticletitle{Program synthesis with large language models}.
\newblock \bibinfo{journal}{\emph{arXiv preprint arXiv:2108.07732}} (\bibinfo{year}{2021}).
\newblock


\bibitem[Bai et~al\mbox{.}(2023)]%
        {qwen}
\bibfield{author}{\bibinfo{person}{Jinze Bai}, \bibinfo{person}{Shuai Bai}, \bibinfo{person}{Yunfei Chu}, \bibinfo{person}{Zeyu Cui}, \bibinfo{person}{Kai Dang}, \bibinfo{person}{Xiaodong Deng}, \bibinfo{person}{Yang Fan}, \bibinfo{person}{Wenbin Ge}, \bibinfo{person}{Yu Han}, \bibinfo{person}{Fei Huang}, \bibinfo{person}{Binyuan Hui}, \bibinfo{person}{Luo Ji}, \bibinfo{person}{Mei Li}, \bibinfo{person}{Junyang Lin}, \bibinfo{person}{Runji Lin}, \bibinfo{person}{Dayiheng Liu}, \bibinfo{person}{Gao Liu}, \bibinfo{person}{Chengqiang Lu}, \bibinfo{person}{Keming Lu}, \bibinfo{person}{Jianxin Ma}, \bibinfo{person}{Rui Men}, \bibinfo{person}{Xingzhang Ren}, \bibinfo{person}{Xuancheng Ren}, \bibinfo{person}{Chuanqi Tan}, \bibinfo{person}{Sinan Tan}, \bibinfo{person}{Jianhong Tu}, \bibinfo{person}{Peng Wang}, \bibinfo{person}{Shijie Wang}, \bibinfo{person}{Wei Wang}, \bibinfo{person}{Shengguang Wu}, \bibinfo{person}{Benfeng Xu}, \bibinfo{person}{Jin Xu}, \bibinfo{person}{An Yang}, \bibinfo{person}{Hao Yang},
  \bibinfo{person}{Jian Yang}, \bibinfo{person}{Shusheng Yang}, \bibinfo{person}{Yang Yao}, \bibinfo{person}{Bowen Yu}, \bibinfo{person}{Hongyi Yuan}, \bibinfo{person}{Zheng Yuan}, \bibinfo{person}{Jianwei Zhang}, \bibinfo{person}{Xingxuan Zhang}, \bibinfo{person}{Yichang Zhang}, \bibinfo{person}{Zhenru Zhang}, \bibinfo{person}{Chang Zhou}, \bibinfo{person}{Jingren Zhou}, \bibinfo{person}{Xiaohuan Zhou}, {and} \bibinfo{person}{Tianhang Zhu}.} \bibinfo{year}{2023}\natexlab{}.
\newblock \bibinfo{title}{Qwen Technical Report}.
\newblock
\newblock
\showeprint[arxiv]{2309.16609}~[cs.CL]
\urldef\tempurl%
\url{https://arxiv.org/abs/2309.16609}
\showURL{%
\tempurl}


\bibitem[Bogner and Merkel(2022)]%
        {bogner2022type}
\bibfield{author}{\bibinfo{person}{Justus Bogner} {and} \bibinfo{person}{Manuel Merkel}.} \bibinfo{year}{2022}\natexlab{}.
\newblock \showarticletitle{To type or not to type? a systematic comparison of the software quality of javascript and typescript applications on github}. In \bibinfo{booktitle}{\emph{Proceedings of the 19th International Conference on Mining Software Repositories}}. \bibinfo{pages}{658--669}.
\newblock


\bibitem[Cassano et~al\mbox{.}(2024)]%
        {cassano2024knowledge}
\bibfield{author}{\bibinfo{person}{Federico Cassano}, \bibinfo{person}{John Gouwar}, \bibinfo{person}{Francesca Lucchetti}, \bibinfo{person}{Claire Schlesinger}, \bibinfo{person}{Anders Freeman}, \bibinfo{person}{Carolyn~Jane Anderson}, \bibinfo{person}{Molly~Q Feldman}, \bibinfo{person}{Michael Greenberg}, \bibinfo{person}{Abhinav Jangda}, {and} \bibinfo{person}{Arjun Guha}.} \bibinfo{year}{2024}\natexlab{}.
\newblock \showarticletitle{Knowledge transfer from high-resource to low-resource programming languages for code llms}.
\newblock \bibinfo{journal}{\emph{Proceedings of the ACM on Programming Languages}} \bibinfo{volume}{8}, \bibinfo{number}{OOPSLA2} (\bibinfo{year}{2024}), \bibinfo{pages}{677--708}.
\newblock


\bibitem[Cassano et~al\mbox{.}(2023)]%
        {multipl}
\bibfield{author}{\bibinfo{person}{Federico Cassano}, \bibinfo{person}{John Gouwar}, \bibinfo{person}{Daniel Nguyen}, \bibinfo{person}{Sydney Nguyen}, \bibinfo{person}{Luna Phipps-Costin}, \bibinfo{person}{Donald Pinckney}, \bibinfo{person}{Ming-Ho Yee}, \bibinfo{person}{Yangtian Zi}, \bibinfo{person}{Carolyn~Jane Anderson}, \bibinfo{person}{Molly~Q Feldman}, {et~al\mbox{.}}} \bibinfo{year}{2023}\natexlab{}.
\newblock \showarticletitle{MultiPL-E: a scalable and polyglot approach to benchmarking neural code generation}.
\newblock \bibinfo{journal}{\emph{IEEE Transactions on Software Engineering}} \bibinfo{volume}{49}, \bibinfo{number}{7} (\bibinfo{year}{2023}), \bibinfo{pages}{3675--3691}.
\newblock


\bibitem[Chen et~al\mbox{.}(2022)]%
        {codet}
\bibfield{author}{\bibinfo{person}{Bei Chen}, \bibinfo{person}{Fengji Zhang}, \bibinfo{person}{Anh Nguyen}, \bibinfo{person}{Daoguang Zan}, \bibinfo{person}{Zeqi Lin}, \bibinfo{person}{Jian-Guang Lou}, {and} \bibinfo{person}{Weizhu Chen}.} \bibinfo{year}{2022}\natexlab{}.
\newblock \showarticletitle{Codet: Code generation with generated tests}.
\newblock \bibinfo{journal}{\emph{arXiv preprint arXiv:2207.10397}} (\bibinfo{year}{2022}).
\newblock


\bibitem[Chen et~al\mbox{.}(2024a)]%
        {chen2024identifying}
\bibfield{author}{\bibinfo{person}{Jiachi Chen}, \bibinfo{person}{Chong Chen}, \bibinfo{person}{Jiang Hu}, \bibinfo{person}{John Grundy}, \bibinfo{person}{Yanlin Wang}, \bibinfo{person}{Ting Chen}, {and} \bibinfo{person}{Zibin Zheng}.} \bibinfo{year}{2024}\natexlab{a}.
\newblock \showarticletitle{Identifying smart contract security issues in code snippets from stack overflow}. In \bibinfo{booktitle}{\emph{Proceedings of the 33rd ACM SIGSOFT International Symposium on Software Testing and Analysis}}. \bibinfo{pages}{1198--1210}.
\newblock


\bibitem[Chen et~al\mbox{.}(2024b)]%
        {chen2024rmcbench}
\bibfield{author}{\bibinfo{person}{Jiachi Chen}, \bibinfo{person}{Qingyuan Zhong}, \bibinfo{person}{Yanlin Wang}, \bibinfo{person}{Kaiwen Ning}, \bibinfo{person}{Yongkun Liu}, \bibinfo{person}{Zenan Xu}, \bibinfo{person}{Zhe Zhao}, \bibinfo{person}{Ting Chen}, {and} \bibinfo{person}{Zibin Zheng}.} \bibinfo{year}{2024}\natexlab{b}.
\newblock \showarticletitle{RMCBench: Benchmarking Large Language Models' Resistance to Malicious Code}. In \bibinfo{booktitle}{\emph{Proceedings of the 39th IEEE/ACM International Conference on Automated Software Engineering}}. \bibinfo{pages}{995--1006}.
\newblock


\bibitem[Chen et~al\mbox{.}(2021)]%
        {humaneval}
\bibfield{author}{\bibinfo{person}{Mark Chen}, \bibinfo{person}{Jerry Tworek}, \bibinfo{person}{Heewoo Jun}, \bibinfo{person}{Qiming Yuan}, \bibinfo{person}{Henrique Ponde De~Oliveira Pinto}, \bibinfo{person}{Jared Kaplan}, \bibinfo{person}{Harri Edwards}, \bibinfo{person}{Yuri Burda}, \bibinfo{person}{Nicholas Joseph}, \bibinfo{person}{Greg Brockman}, {et~al\mbox{.}}} \bibinfo{year}{2021}\natexlab{}.
\newblock \showarticletitle{Evaluating large language models trained on code}.
\newblock \bibinfo{journal}{\emph{arXiv preprint arXiv:2107.03374}} (\bibinfo{year}{2021}).
\newblock


\bibitem[Chowdhery et~al\mbox{.}(2023)]%
        {chowdhery2023palm}
\bibfield{author}{\bibinfo{person}{Aakanksha Chowdhery}, \bibinfo{person}{Sharan Narang}, \bibinfo{person}{Jacob Devlin}, \bibinfo{person}{Maarten Bosma}, \bibinfo{person}{Gaurav Mishra}, \bibinfo{person}{Adam Roberts}, \bibinfo{person}{Paul Barham}, \bibinfo{person}{Hyung~Won Chung}, \bibinfo{person}{Charles Sutton}, \bibinfo{person}{Sebastian Gehrmann}, {et~al\mbox{.}}} \bibinfo{year}{2023}\natexlab{}.
\newblock \showarticletitle{Palm: Scaling language modeling with pathways}.
\newblock \bibinfo{journal}{\emph{Journal of Machine Learning Research}} \bibinfo{volume}{24}, \bibinfo{number}{240} (\bibinfo{year}{2023}), \bibinfo{pages}{1--113}.
\newblock


\bibitem[Cobbe et~al\mbox{.}(2021)]%
        {cobbe2021training}
\bibfield{author}{\bibinfo{person}{Karl Cobbe}, \bibinfo{person}{Vineet Kosaraju}, \bibinfo{person}{Mohammad Bavarian}, \bibinfo{person}{Mark Chen}, \bibinfo{person}{Heewoo Jun}, \bibinfo{person}{Lukasz Kaiser}, \bibinfo{person}{Matthias Plappert}, \bibinfo{person}{Jerry Tworek}, \bibinfo{person}{Jacob Hilton}, \bibinfo{person}{Reiichiro Nakano}, {et~al\mbox{.}}} \bibinfo{year}{2021}\natexlab{}.
\newblock \showarticletitle{Training verifiers to solve math word problems}.
\newblock \bibinfo{journal}{\emph{arXiv preprint arXiv:2110.14168}} (\bibinfo{year}{2021}).
\newblock


\bibitem[Croft et~al\mbox{.}(2022)]%
        {croft2022empirical}
\bibfield{author}{\bibinfo{person}{Roland Croft}, \bibinfo{person}{Yongzheng Xie}, \bibinfo{person}{Mansooreh Zahedi}, \bibinfo{person}{M~Ali Babar}, {and} \bibinfo{person}{Christoph Treude}.} \bibinfo{year}{2022}\natexlab{}.
\newblock \showarticletitle{An empirical study of developers’ discussions about security challenges of different programming languages}.
\newblock \bibinfo{journal}{\emph{Empirical Software Engineering}}  \bibinfo{volume}{27} (\bibinfo{year}{2022}), \bibinfo{pages}{1--52}.
\newblock


\bibitem[Cursor({[n.\,d.]})]%
        {Cursor}
Cursor \bibinfo{year}{[n.\,d.]}\natexlab{}.
\newblock
\newblock
\urldef\tempurl%
\url{https://www.cursor.com/}
\showURL{%
\tempurl}


\bibitem[Dehaerne et~al\mbox{.}(2022)]%
        {codegeneration1}
\bibfield{author}{\bibinfo{person}{Enrique Dehaerne}, \bibinfo{person}{Bappaditya Dey}, \bibinfo{person}{Sandip Halder}, \bibinfo{person}{Stefan De~Gendt}, {and} \bibinfo{person}{Wannes Meert}.} \bibinfo{year}{2022}\natexlab{}.
\newblock \showarticletitle{Code generation using machine learning: A systematic review}.
\newblock \bibinfo{journal}{\emph{Ieee Access}}  \bibinfo{volume}{10} (\bibinfo{year}{2022}), \bibinfo{pages}{82434--82455}.
\newblock


\bibitem[Dev({[n.\,d.]})]%
        {Dev}
Dev \bibinfo{year}{[n.\,d.]}\natexlab{}.
\newblock
\newblock
\urldef\tempurl%
\url{https://dev.to/}
\showURL{%
\tempurl}


\bibitem[Ding et~al\mbox{.}(2024)]%
        {crosscodeeval}
\bibfield{author}{\bibinfo{person}{Yangruibo Ding}, \bibinfo{person}{Zijian Wang}, \bibinfo{person}{Wasi Ahmad}, \bibinfo{person}{Hantian Ding}, \bibinfo{person}{Ming Tan}, \bibinfo{person}{Nihal Jain}, \bibinfo{person}{Murali~Krishna Ramanathan}, \bibinfo{person}{Ramesh Nallapati}, \bibinfo{person}{Parminder Bhatia}, \bibinfo{person}{Dan Roth}, {et~al\mbox{.}}} \bibinfo{year}{2024}\natexlab{}.
\newblock \showarticletitle{Crosscodeeval: A diverse and multilingual benchmark for cross-file code completion}.
\newblock \bibinfo{journal}{\emph{Advances in Neural Information Processing Systems}}  \bibinfo{volume}{36} (\bibinfo{year}{2024}).
\newblock


\bibitem[Ding et~al\mbox{.}(2022)]%
        {ding2022cocomic}
\bibfield{author}{\bibinfo{person}{Yangruibo Ding}, \bibinfo{person}{Zijian Wang}, \bibinfo{person}{Wasi~Uddin Ahmad}, \bibinfo{person}{Murali~Krishna Ramanathan}, \bibinfo{person}{Ramesh Nallapati}, \bibinfo{person}{Parminder Bhatia}, \bibinfo{person}{Dan Roth}, {and} \bibinfo{person}{Bing Xiang}.} \bibinfo{year}{2022}\natexlab{}.
\newblock \showarticletitle{Cocomic: Code completion by jointly modeling in-file and cross-file context}.
\newblock \bibinfo{journal}{\emph{arXiv preprint arXiv:2212.10007}} (\bibinfo{year}{2022}).
\newblock


\bibitem[Dinh and Wang(2020)]%
        {dinh2020modern}
\bibfield{author}{\bibinfo{person}{Duong Dinh} {and} \bibinfo{person}{Zhuanyan Wang}.} \bibinfo{year}{2020}\natexlab{}.
\newblock \showarticletitle{Modern front-end web development: how libraries and frameworks transform everything}.
\newblock  (\bibinfo{year}{2020}).
\newblock


\bibitem[Dohmke et~al\mbox{.}(2023)]%
        {githubcopilot}
\bibfield{author}{\bibinfo{person}{Thomas Dohmke}, \bibinfo{person}{Marco Iansiti}, {and} \bibinfo{person}{Greg Richards}.} \bibinfo{year}{2023}\natexlab{}.
\newblock \showarticletitle{Sea change in software development: Economic and productivity analysis of the ai-powered developer lifecycle}.
\newblock \bibinfo{journal}{\emph{arXiv preprint arXiv:2306.15033}} (\bibinfo{year}{2023}).
\newblock


\bibitem[DomainCodeBench({[n.\,d.]})]%
        {DomainCodeBench}
DomainCodeBench \bibinfo{year}{[n.\,d.]}\natexlab{}.
\newblock
\newblock
\urldef\tempurl%
\url{https://github.com/DeepSoftwareAnalytics/DomainCodeBench}
\showURL{%
\tempurl}


\bibitem[Dong et~al\mbox{.}(2024)]%
        {dong2024self}
\bibfield{author}{\bibinfo{person}{Yihong Dong}, \bibinfo{person}{Xue Jiang}, \bibinfo{person}{Zhi Jin}, {and} \bibinfo{person}{Ge Li}.} \bibinfo{year}{2024}\natexlab{}.
\newblock \showarticletitle{Self-collaboration code generation via chatgpt}.
\newblock \bibinfo{journal}{\emph{ACM Transactions on Software Engineering and Methodology}} \bibinfo{volume}{33}, \bibinfo{number}{7} (\bibinfo{year}{2024}), \bibinfo{pages}{1--38}.
\newblock


\bibitem[Du et~al\mbox{.}(2023)]%
        {classeval}
\bibfield{author}{\bibinfo{person}{Xueying Du}, \bibinfo{person}{Mingwei Liu}, \bibinfo{person}{Kaixin Wang}, \bibinfo{person}{Hanlin Wang}, \bibinfo{person}{Junwei Liu}, \bibinfo{person}{Yixuan Chen}, \bibinfo{person}{Jiayi Feng}, \bibinfo{person}{Chaofeng Sha}, \bibinfo{person}{Xin Peng}, {and} \bibinfo{person}{Yiling Lou}.} \bibinfo{year}{2023}\natexlab{}.
\newblock \showarticletitle{Classeval: A manually-crafted benchmark for evaluating llms on class-level code generation}.
\newblock \bibinfo{journal}{\emph{arXiv preprint arXiv:2308.01861}} (\bibinfo{year}{2023}).
\newblock


\bibitem[Fan et~al\mbox{.}(2024)]%
        {fan2024exploring}
\bibfield{author}{\bibinfo{person}{Lishui Fan}, \bibinfo{person}{Jiakun Liu}, \bibinfo{person}{Zhongxin Liu}, \bibinfo{person}{David Lo}, \bibinfo{person}{Xin Xia}, {and} \bibinfo{person}{Shanping Li}.} \bibinfo{year}{2024}\natexlab{}.
\newblock \showarticletitle{Exploring the capabilities of llms for code change related tasks}.
\newblock \bibinfo{journal}{\emph{arXiv preprint arXiv:2407.02824}} (\bibinfo{year}{2024}).
\newblock


\bibitem[Feng et~al\mbox{.}(2024)]%
        {ComplexCodeEval}
\bibfield{author}{\bibinfo{person}{Jia Feng}, \bibinfo{person}{Jiachen Liu}, \bibinfo{person}{Cuiyun Gao}, \bibinfo{person}{Chun~Yong Chong}, \bibinfo{person}{Chaozheng Wang}, \bibinfo{person}{Shan Gao}, {and} \bibinfo{person}{Xin Xia}.} \bibinfo{year}{2024}\natexlab{}.
\newblock \showarticletitle{ComplexCodeEval: A Benchmark for Evaluating Large Code Models on More Complex Code} \emph{(\bibinfo{series}{ASE '24})}. \bibinfo{publisher}{Association for Computing Machinery}, \bibinfo{address}{New York, NY, USA}, \bibinfo{pages}{1895–1906}.
\newblock
\showISBNx{9798400712487}
\urldef\tempurl%
\url{https://doi.org/10.1145/3691620.3695552}
\showDOI{\tempurl}


\bibitem[Fried et~al\mbox{.}(2022)]%
        {fried2022incoder}
\bibfield{author}{\bibinfo{person}{Daniel Fried}, \bibinfo{person}{Armen Aghajanyan}, \bibinfo{person}{Jessy Lin}, \bibinfo{person}{Sida Wang}, \bibinfo{person}{Eric Wallace}, \bibinfo{person}{Freda Shi}, \bibinfo{person}{Ruiqi Zhong}, \bibinfo{person}{Wen-tau Yih}, \bibinfo{person}{Luke Zettlemoyer}, {and} \bibinfo{person}{Mike Lewis}.} \bibinfo{year}{2022}\natexlab{}.
\newblock \showarticletitle{Incoder: A generative model for code infilling and synthesis}.
\newblock \bibinfo{journal}{\emph{arXiv preprint arXiv:2204.05999}} (\bibinfo{year}{2022}).
\newblock


\bibitem[GitHub Blog({[n.\,d.]})]%
        {githubblog}
GitHub Blog \bibinfo{year}{[n.\,d.]}\natexlab{}.
\newblock
\newblock
\urldef\tempurl%
\url{https://github.blog/}
\showURL{%
\tempurl}


\bibitem[Gu et~al\mbox{.}(2024a)]%
        {domcoder}
\bibfield{author}{\bibinfo{person}{Xiaodong Gu}, \bibinfo{person}{Meng Chen}, \bibinfo{person}{Yalan Lin}, \bibinfo{person}{Yuhan Hu}, \bibinfo{person}{Hongyu Zhang}, \bibinfo{person}{Chengcheng Wan}, \bibinfo{person}{Zhao Wei}, \bibinfo{person}{Yong Xu}, {and} \bibinfo{person}{Juhong Wang}.} \bibinfo{year}{2024}\natexlab{a}.
\newblock \showarticletitle{On the effectiveness of large language models in domain-specific code generation}.
\newblock \bibinfo{journal}{\emph{ACM Transactions on Software Engineering and Methodology}} (\bibinfo{year}{2024}).
\newblock


\bibitem[Gu et~al\mbox{.}(2024b)]%
        {gu2024effectiveness}
\bibfield{author}{\bibinfo{person}{Xiaodong Gu}, \bibinfo{person}{Meng Chen}, \bibinfo{person}{Yalan Lin}, \bibinfo{person}{Yuhan Hu}, \bibinfo{person}{Hongyu Zhang}, \bibinfo{person}{Chengcheng Wan}, \bibinfo{person}{Zhao Wei}, \bibinfo{person}{Yong Xu}, {and} \bibinfo{person}{Juhong Wang}.} \bibinfo{year}{2024}\natexlab{b}.
\newblock \showarticletitle{On the effectiveness of large language models in domain-specific code generation}.
\newblock \bibinfo{journal}{\emph{ACM Transactions on Software Engineering and Methodology}} (\bibinfo{year}{2024}).
\newblock


\bibitem[Guo et~al\mbox{.}(2024)]%
        {guo2024stop}
\bibfield{author}{\bibinfo{person}{Lianghong Guo}, \bibinfo{person}{Yanlin Wang}, \bibinfo{person}{Ensheng Shi}, \bibinfo{person}{Wanjun Zhong}, \bibinfo{person}{Hongyu Zhang}, \bibinfo{person}{Jiachi Chen}, \bibinfo{person}{Ruikai Zhang}, \bibinfo{person}{Yuchi Ma}, {and} \bibinfo{person}{Zibin Zheng}.} \bibinfo{year}{2024}\natexlab{}.
\newblock \showarticletitle{When to stop? towards efficient code generation in llms with excess token prevention}. In \bibinfo{booktitle}{\emph{Proceedings of the 33rd ACM SIGSOFT International Symposium on Software Testing and Analysis}}. \bibinfo{pages}{1073--1085}.
\newblock


\bibitem[Hacker News({[n.\,d.]})]%
        {HackerNews}
Hacker News \bibinfo{year}{[n.\,d.]}\natexlab{}.
\newblock
\newblock
\urldef\tempurl%
\url{https://news.ycombinator.com/}
\showURL{%
\tempurl}


\bibitem[Hao et~al\mbox{.}(2022)]%
        {aixbench}
\bibfield{author}{\bibinfo{person}{Yiyang Hao}, \bibinfo{person}{Ge Li}, \bibinfo{person}{Yongqiang Liu}, \bibinfo{person}{Xiaowei Miao}, \bibinfo{person}{He Zong}, \bibinfo{person}{Siyuan Jiang}, \bibinfo{person}{Yang Liu}, {and} \bibinfo{person}{He Wei}.} \bibinfo{year}{2022}\natexlab{}.
\newblock \showarticletitle{Aixbench: A code generation benchmark dataset}.
\newblock \bibinfo{journal}{\emph{arXiv preprint arXiv:2206.13179}} (\bibinfo{year}{2022}).
\newblock


\bibitem[Hong et~al\mbox{.}(2024)]%
        {metagpt}
\bibfield{author}{\bibinfo{person}{Sirui Hong}, \bibinfo{person}{Mingchen Zhuge}, \bibinfo{person}{Jonathan Chen}, \bibinfo{person}{Xiawu Zheng}, \bibinfo{person}{Yuheng Cheng}, \bibinfo{person}{Jinlin Wang}, \bibinfo{person}{Ceyao Zhang}, \bibinfo{person}{Zili Wang}, \bibinfo{person}{Steven Ka~Shing Yau}, \bibinfo{person}{Zijuan Lin}, \bibinfo{person}{Liyang Zhou}, \bibinfo{person}{Chenyu Ran}, \bibinfo{person}{Lingfeng Xiao}, \bibinfo{person}{Chenglin Wu}, {and} \bibinfo{person}{J{\"u}rgen Schmidhuber}.} \bibinfo{year}{2024}\natexlab{}.
\newblock \showarticletitle{Meta{GPT}: Meta Programming for Multi-Agent Collaborative Framework}. In \bibinfo{booktitle}{\emph{The Twelfth International Conference on Learning Representations}}.
\newblock


\bibitem[Iyer et~al\mbox{.}(2018)]%
        {concode}
\bibfield{author}{\bibinfo{person}{Srinivasan Iyer}, \bibinfo{person}{Ioannis Konstas}, \bibinfo{person}{Alvin Cheung}, {and} \bibinfo{person}{Luke Zettlemoyer}.} \bibinfo{year}{2018}\natexlab{}.
\newblock \bibinfo{title}{Mapping Language to Code in Programmatic Context}.
\newblock
\newblock
\showeprint[arxiv]{1808.09588}~[cs.CL]
\urldef\tempurl%
\url{https://arxiv.org/abs/1808.09588}
\showURL{%
\tempurl}


\bibitem[Jelodar et~al\mbox{.}(2019)]%
        {jelodar2019latent}
\bibfield{author}{\bibinfo{person}{Hamed Jelodar}, \bibinfo{person}{Yongli Wang}, \bibinfo{person}{Chi Yuan}, \bibinfo{person}{Xia Feng}, \bibinfo{person}{Xiahui Jiang}, \bibinfo{person}{Yanchao Li}, {and} \bibinfo{person}{Liang Zhao}.} \bibinfo{year}{2019}\natexlab{}.
\newblock \showarticletitle{Latent Dirichlet allocation (LDA) and topic modeling: models, applications, a survey}.
\newblock \bibinfo{journal}{\emph{Multimedia tools and applications}}  \bibinfo{volume}{78} (\bibinfo{year}{2019}), \bibinfo{pages}{15169--15211}.
\newblock


\bibitem[Jiang et~al\mbox{.}(2023)]%
        {jiang2023selfevolve}
\bibfield{author}{\bibinfo{person}{Shuyang Jiang}, \bibinfo{person}{Yuhao Wang}, {and} \bibinfo{person}{Yu Wang}.} \bibinfo{year}{2023}\natexlab{}.
\newblock \showarticletitle{Selfevolve: A code evolution framework via large language models}.
\newblock \bibinfo{journal}{\emph{arXiv preprint arXiv:2306.02907}} (\bibinfo{year}{2023}).
\newblock


\bibitem[Kernycky et~al\mbox{.}(2024)]%
        {kernycky2024evaluating}
\bibfield{author}{\bibinfo{person}{Andrew Kernycky}, \bibinfo{person}{David Coleman}, \bibinfo{person}{Christopher Spence}, {and} \bibinfo{person}{Udayan Das}.} \bibinfo{year}{2024}\natexlab{}.
\newblock \showarticletitle{Evaluating the Performance of LLMs on Technical Language Processing Tasks}. In \bibinfo{booktitle}{\emph{International Conference on Human-Computer Interaction}}. Springer, \bibinfo{pages}{75--85}.
\newblock


\bibitem[Kolasani(2023)]%
        {kolasani2023optimizing}
\bibfield{author}{\bibinfo{person}{Saydulu Kolasani}.} \bibinfo{year}{2023}\natexlab{}.
\newblock \showarticletitle{Optimizing natural language processing, large language models (LLMs) for efficient customer service, and hyper-personalization to enable sustainable growth and revenue}.
\newblock \bibinfo{journal}{\emph{Transactions on Latest Trends in Artificial Intelligence}} \bibinfo{volume}{4}, \bibinfo{number}{4} (\bibinfo{year}{2023}).
\newblock


\bibitem[Lai et~al\mbox{.}(2022)]%
        {DS-1000}
\bibfield{author}{\bibinfo{person}{Yuhang Lai}, \bibinfo{person}{Chengxi Li}, \bibinfo{person}{Yiming Wang}, \bibinfo{person}{Tianyi Zhang}, \bibinfo{person}{Ruiqi Zhong}, \bibinfo{person}{Luke Zettlemoyer}, \bibinfo{person}{Scott~Wen tau Yih}, \bibinfo{person}{Daniel Fried}, \bibinfo{person}{Sida Wang}, {and} \bibinfo{person}{Tao Yu}.} \bibinfo{year}{2022}\natexlab{}.
\newblock \bibinfo{title}{DS-1000: A Natural and Reliable Benchmark for Data Science Code Generation}.
\newblock
\newblock
\showeprint[arxiv]{2211.11501}~[cs.SE]
\urldef\tempurl%
\url{https://arxiv.org/abs/2211.11501}
\showURL{%
\tempurl}


\bibitem[Le et~al\mbox{.}(2022)]%
        {le2022coderl}
\bibfield{author}{\bibinfo{person}{Hung Le}, \bibinfo{person}{Yue Wang}, \bibinfo{person}{Akhilesh~Deepak Gotmare}, \bibinfo{person}{Silvio Savarese}, {and} \bibinfo{person}{Steven Chu~Hong Hoi}.} \bibinfo{year}{2022}\natexlab{}.
\newblock \showarticletitle{Coderl: Mastering code generation through pretrained models and deep reinforcement learning}.
\newblock \bibinfo{journal}{\emph{Advances in Neural Information Processing Systems}}  \bibinfo{volume}{35} (\bibinfo{year}{2022}), \bibinfo{pages}{21314--21328}.
\newblock


\bibitem[Li et~al\mbox{.}(2024a)]%
        {evocodebench}
\bibfield{author}{\bibinfo{person}{Jia Li}, \bibinfo{person}{Ge Li}, \bibinfo{person}{Xuanming Zhang}, \bibinfo{person}{Yunfei Zhao}, \bibinfo{person}{Yihong Dong}, \bibinfo{person}{Zhi Jin}, \bibinfo{person}{Binhua Li}, \bibinfo{person}{Fei Huang}, {and} \bibinfo{person}{Yongbin Li}.} \bibinfo{year}{2024}\natexlab{a}.
\newblock \showarticletitle{EvoCodeBench: An Evolving Code Generation Benchmark with Domain-Specific Evaluations}. In \bibinfo{booktitle}{\emph{The Thirty-eight Conference on Neural Information Processing Systems Datasets and Benchmarks Track}}.
\newblock
\urldef\tempurl%
\url{https://openreview.net/forum?id=kvjbFVHpny}
\showURL{%
\tempurl}


\bibitem[Li et~al\mbox{.}(2023)]%
        {starcoder}
\bibfield{author}{\bibinfo{person}{Raymond Li}, \bibinfo{person}{Loubna~Ben Allal}, \bibinfo{person}{Yangtian Zi}, \bibinfo{person}{Niklas Muennighoff}, \bibinfo{person}{Denis Kocetkov}, \bibinfo{person}{Chenghao Mou}, \bibinfo{person}{Marc Marone}, \bibinfo{person}{Christopher Akiki}, \bibinfo{person}{Jia Li}, \bibinfo{person}{Jenny Chim}, {et~al\mbox{.}}} \bibinfo{year}{2023}\natexlab{}.
\newblock \showarticletitle{Starcoder: may the source be with you!}
\newblock \bibinfo{journal}{\emph{arXiv preprint arXiv:2305.06161}} (\bibinfo{year}{2023}).
\newblock


\bibitem[Li et~al\mbox{.}(2022)]%
        {li2022competition}
\bibfield{author}{\bibinfo{person}{Yujia Li}, \bibinfo{person}{David Choi}, \bibinfo{person}{Junyoung Chung}, \bibinfo{person}{Nate Kushman}, \bibinfo{person}{Julian Schrittwieser}, \bibinfo{person}{R{\'e}mi Leblond}, \bibinfo{person}{Tom Eccles}, \bibinfo{person}{James Keeling}, \bibinfo{person}{Felix Gimeno}, \bibinfo{person}{Agustin Dal~Lago}, {et~al\mbox{.}}} \bibinfo{year}{2022}\natexlab{}.
\newblock \showarticletitle{Competition-level code generation with alphacode}.
\newblock \bibinfo{journal}{\emph{Science}} \bibinfo{volume}{378}, \bibinfo{number}{6624} (\bibinfo{year}{2022}), \bibinfo{pages}{1092--1097}.
\newblock


\bibitem[Li et~al\mbox{.}(2024b)]%
        {li2024enhancing}
\bibfield{author}{\bibinfo{person}{Yichen Li}, \bibinfo{person}{Yun Peng}, \bibinfo{person}{Yintong Huo}, {and} \bibinfo{person}{Michael~R Lyu}.} \bibinfo{year}{2024}\natexlab{b}.
\newblock \showarticletitle{Enhancing llm-based coding tools through native integration of ide-derived static context}. In \bibinfo{booktitle}{\emph{Proceedings of the 1st International Workshop on Large Language Models for Code}}. \bibinfo{pages}{70--74}.
\newblock


\bibitem[Liu et~al\mbox{.}(2023)]%
        {codegen4libs}
\bibfield{author}{\bibinfo{person}{Mingwei Liu}, \bibinfo{person}{Tianyong Yang}, \bibinfo{person}{Yiling Lou}, \bibinfo{person}{Xueying Du}, \bibinfo{person}{Ying Wang}, {and} \bibinfo{person}{Xin Peng}.} \bibinfo{year}{2023}\natexlab{}.
\newblock \showarticletitle{Codegen4libs: A two-stage approach for library-oriented code generation}. In \bibinfo{booktitle}{\emph{2023 38th IEEE/ACM International Conference on Automated Software Engineering (ASE)}}. IEEE, \bibinfo{pages}{434--445}.
\newblock


\bibitem[Liu et~al\mbox{.}(2024)]%
        {repobench}
\bibfield{author}{\bibinfo{person}{Tianyang Liu}, \bibinfo{person}{Canwen Xu}, {and} \bibinfo{person}{Julian McAuley}.} \bibinfo{year}{2024}\natexlab{}.
\newblock \showarticletitle{RepoBench: Benchmarking Repository-Level Code Auto-Completion Systems}. In \bibinfo{booktitle}{\emph{The Twelfth International Conference on Learning Representations}}.
\newblock


\bibitem[Lozhkov et~al\mbox{.}(2024)]%
        {starcoder2}
\bibfield{author}{\bibinfo{person}{Anton Lozhkov}, \bibinfo{person}{Raymond Li}, \bibinfo{person}{Loubna~Ben Allal}, \bibinfo{person}{Federico Cassano}, \bibinfo{person}{Joel Lamy-Poirier}, \bibinfo{person}{Nouamane Tazi}, \bibinfo{person}{Ao Tang}, \bibinfo{person}{Dmytro Pykhtar}, \bibinfo{person}{Jiawei Liu}, \bibinfo{person}{Yuxiang Wei}, {et~al\mbox{.}}} \bibinfo{year}{2024}\natexlab{}.
\newblock \showarticletitle{Starcoder 2 and the stack v2: The next generation}.
\newblock \bibinfo{journal}{\emph{arXiv preprint arXiv:2402.19173}} (\bibinfo{year}{2024}).
\newblock


\bibitem[Luo et~al\mbox{.}(2023)]%
        {wizardcoder}
\bibfield{author}{\bibinfo{person}{Ziyang Luo}, \bibinfo{person}{Can Xu}, \bibinfo{person}{Pu Zhao}, \bibinfo{person}{Qingfeng Sun}, \bibinfo{person}{Xiubo Geng}, \bibinfo{person}{Wenxiang Hu}, \bibinfo{person}{Chongyang Tao}, \bibinfo{person}{Jing Ma}, \bibinfo{person}{Qingwei Lin}, {and} \bibinfo{person}{Daxin Jiang}.} \bibinfo{year}{2023}\natexlab{}.
\newblock \showarticletitle{Wizardcoder: Empowering code large language models with evol-instruct}.
\newblock \bibinfo{journal}{\emph{arXiv preprint arXiv:2306.08568}} (\bibinfo{year}{2023}).
\newblock


\bibitem[Maier et~al\mbox{.}(2021)]%
        {maier2021applying}
\bibfield{author}{\bibinfo{person}{Daniel Maier}, \bibinfo{person}{Annie Waldherr}, \bibinfo{person}{Peter Miltner}, \bibinfo{person}{Gregor Wiedemann}, \bibinfo{person}{Andreas Niekler}, \bibinfo{person}{Alexa Keinert}, \bibinfo{person}{Barbara Pfetsch}, \bibinfo{person}{Gerhard Heyer}, \bibinfo{person}{Ueli Reber}, \bibinfo{person}{Thomas H{\"a}ussler}, {et~al\mbox{.}}} \bibinfo{year}{2021}\natexlab{}.
\newblock \showarticletitle{Applying LDA topic modeling in communication research: Toward a valid and reliable methodology}.
\newblock In \bibinfo{booktitle}{\emph{Computational methods for communication science}}. \bibinfo{publisher}{Routledge}, \bibinfo{pages}{13--38}.
\newblock


\bibitem[Medium({[n.\,d.]})]%
        {medium}
Medium \bibinfo{year}{[n.\,d.]}\natexlab{}.
\newblock
\newblock
\urldef\tempurl%
\url{https://medium.com/}
\showURL{%
\tempurl}


\bibitem[Muennighoff et~al\mbox{.}(2023)]%
        {muennighoff2023octopack}
\bibfield{author}{\bibinfo{person}{Niklas Muennighoff}, \bibinfo{person}{Qian Liu}, \bibinfo{person}{Armel Zebaze}, \bibinfo{person}{Qinkai Zheng}, \bibinfo{person}{Binyuan Hui}, \bibinfo{person}{Terry~Yue Zhuo}, \bibinfo{person}{Swayam Singh}, \bibinfo{person}{Xiangru Tang}, \bibinfo{person}{Leandro Von~Werra}, {and} \bibinfo{person}{Shayne Longpre}.} \bibinfo{year}{2023}\natexlab{}.
\newblock \showarticletitle{Octopack: Instruction tuning code large language models}.
\newblock \bibinfo{journal}{\emph{arXiv preprint arXiv:2308.07124}} (\bibinfo{year}{2023}).
\newblock


\bibitem[Nejjar et~al\mbox{.}(2023)]%
        {nejjar2023llms}
\bibfield{author}{\bibinfo{person}{Mohamed Nejjar}, \bibinfo{person}{Luca Zacharias}, \bibinfo{person}{Fabian Stiehle}, {and} \bibinfo{person}{Ingo Weber}.} \bibinfo{year}{2023}\natexlab{}.
\newblock \showarticletitle{LLMs for science: Usage for code generation and data analysis}.
\newblock \bibinfo{journal}{\emph{Journal of Software: Evolution and Process}} (\bibinfo{year}{2023}), \bibinfo{pages}{e2723}.
\newblock


\bibitem[Nijkamp et~al\mbox{.}(2022)]%
        {nijkamp2022codegen}
\bibfield{author}{\bibinfo{person}{Erik Nijkamp}, \bibinfo{person}{Bo Pang}, \bibinfo{person}{Hiroaki Hayashi}, \bibinfo{person}{Lifu Tu}, \bibinfo{person}{Huan Wang}, \bibinfo{person}{Yingbo Zhou}, \bibinfo{person}{Silvio Savarese}, {and} \bibinfo{person}{Caiming Xiong}.} \bibinfo{year}{2022}\natexlab{}.
\newblock \showarticletitle{Codegen: An open large language model for code with multi-turn program synthesis}.
\newblock \bibinfo{journal}{\emph{arXiv preprint arXiv:2203.13474}} (\bibinfo{year}{2022}).
\newblock


\bibitem[Pang et~al\mbox{.}(2024)]%
        {pang2024ai2apps}
\bibfield{author}{\bibinfo{person}{Xin Pang}, \bibinfo{person}{Zhucong Li}, \bibinfo{person}{Jiaxiang Chen}, \bibinfo{person}{Yuan Cheng}, \bibinfo{person}{Yinghui Xu}, {and} \bibinfo{person}{Yuan Qi}.} \bibinfo{year}{2024}\natexlab{}.
\newblock \showarticletitle{AI2Apps: A Visual IDE for Building LLM-based AI Agent Applications}.
\newblock \bibinfo{journal}{\emph{arXiv preprint arXiv:2404.04902}} (\bibinfo{year}{2024}).
\newblock


\bibitem[Papineni et~al\mbox{.}(2002)]%
        {bleu}
\bibfield{author}{\bibinfo{person}{Kishore Papineni}, \bibinfo{person}{Salim Roukos}, \bibinfo{person}{Todd Ward}, {and} \bibinfo{person}{Wei-Jing Zhu}.} \bibinfo{year}{2002}\natexlab{}.
\newblock \showarticletitle{BLEU: a method for automatic evaluation of machine translation}. In \bibinfo{booktitle}{\emph{Proceedings of the 40th Annual Meeting on Association for Computational Linguistics}} (Philadelphia, Pennsylvania) \emph{(\bibinfo{series}{ACL '02})}. \bibinfo{publisher}{Association for Computational Linguistics}, \bibinfo{address}{USA}, \bibinfo{pages}{311–318}.
\newblock
\urldef\tempurl%
\url{https://doi.org/10.3115/1073083.1073135}
\showDOI{\tempurl}


\bibitem[Qian et~al\mbox{.}(2023)]%
        {chatdev}
\bibfield{author}{\bibinfo{person}{Chen Qian}, \bibinfo{person}{Xin Cong}, \bibinfo{person}{Wei Liu}, \bibinfo{person}{Cheng Yang}, \bibinfo{person}{Weize Chen}, \bibinfo{person}{Yusheng Su}, \bibinfo{person}{Yufan Dang}, \bibinfo{person}{Jiahao Li}, \bibinfo{person}{Juyuan Xu}, \bibinfo{person}{Dahai Li}, \bibinfo{person}{Zhiyuan Liu}, {and} \bibinfo{person}{Maosong Sun}.} \bibinfo{year}{2023}\natexlab{}.
\newblock \bibinfo{title}{Communicative Agents for Software Development}.
\newblock
\newblock
\showeprint[arxiv]{2307.07924}~[cs.SE]


\bibitem[Ren et~al\mbox{.}(2020)]%
        {codebleu}
\bibfield{author}{\bibinfo{person}{Shuo Ren}, \bibinfo{person}{Daya Guo}, \bibinfo{person}{Shuai Lu}, \bibinfo{person}{Long Zhou}, \bibinfo{person}{Shujie Liu}, \bibinfo{person}{Duyu Tang}, \bibinfo{person}{Neel Sundaresan}, \bibinfo{person}{Ming Zhou}, \bibinfo{person}{Ambrosio Blanco}, {and} \bibinfo{person}{Shuai Ma}.} \bibinfo{year}{2020}\natexlab{}.
\newblock \bibinfo{title}{CodeBLEU: a Method for Automatic Evaluation of Code Synthesis}.
\newblock
\newblock
\showeprint[arxiv]{2009.10297}~[cs.SE]
\urldef\tempurl%
\url{https://arxiv.org/abs/2009.10297}
\showURL{%
\tempurl}


\bibitem[Roziere et~al\mbox{.}(2023)]%
        {codellama}
\bibfield{author}{\bibinfo{person}{Baptiste Roziere}, \bibinfo{person}{Jonas Gehring}, \bibinfo{person}{Fabian Gloeckle}, \bibinfo{person}{Sten Sootla}, \bibinfo{person}{Itai Gat}, \bibinfo{person}{Xiaoqing~Ellen Tan}, \bibinfo{person}{Yossi Adi}, \bibinfo{person}{Jingyu Liu}, \bibinfo{person}{Romain Sauvestre}, \bibinfo{person}{Tal Remez}, {et~al\mbox{.}}} \bibinfo{year}{2023}\natexlab{}.
\newblock \showarticletitle{Code llama: Open foundation models for code}.
\newblock \bibinfo{journal}{\emph{arXiv preprint arXiv:2308.12950}} (\bibinfo{year}{2023}).
\newblock


\bibitem[Ryder et~al\mbox{.}(2005)]%
        {ryder2005impact}
\bibfield{author}{\bibinfo{person}{Barbara~G Ryder}, \bibinfo{person}{Mary~Lou Soffa}, {and} \bibinfo{person}{Margaret Burnett}.} \bibinfo{year}{2005}\natexlab{}.
\newblock \showarticletitle{The impact of software engineering research on modern programming languages}.
\newblock \bibinfo{journal}{\emph{ACM Transactions on Software Engineering and Methodology (TOSEM)}} \bibinfo{volume}{14}, \bibinfo{number}{4} (\bibinfo{year}{2005}), \bibinfo{pages}{431--477}.
\newblock


\bibitem[Samek(2008)]%
        {samek2008practical}
\bibfield{author}{\bibinfo{person}{Miro Samek}.} \bibinfo{year}{2008}\natexlab{}.
\newblock \bibinfo{booktitle}{\emph{Practical UML statecharts in C/C++: event-driven programming for embedded systems}}.
\newblock \bibinfo{publisher}{CRC Press}.
\newblock


\bibitem[Shen et~al\mbox{.}(2023)]%
        {pangu}
\bibfield{author}{\bibinfo{person}{Bo Shen}, \bibinfo{person}{Jiaxin Zhang}, \bibinfo{person}{Taihong Chen}, \bibinfo{person}{Daoguang Zan}, \bibinfo{person}{Bing Geng}, \bibinfo{person}{An Fu}, \bibinfo{person}{Muhan Zeng}, \bibinfo{person}{Ailun Yu}, \bibinfo{person}{Jichuan Ji}, \bibinfo{person}{Jingyang Zhao}, {et~al\mbox{.}}} \bibinfo{year}{2023}\natexlab{}.
\newblock \showarticletitle{Pangu-coder2: Boosting large language models for code with ranking feedback}.
\newblock \bibinfo{journal}{\emph{arXiv preprint arXiv:2307.14936}} (\bibinfo{year}{2023}).
\newblock


\bibitem[Shin and Nam(2021)]%
        {codegeneration2}
\bibfield{author}{\bibinfo{person}{Jiho Shin} {and} \bibinfo{person}{Jaechang Nam}.} \bibinfo{year}{2021}\natexlab{}.
\newblock \showarticletitle{A survey of automatic code generation from natural language}.
\newblock \bibinfo{journal}{\emph{Journal of Information Processing Systems}} \bibinfo{volume}{17}, \bibinfo{number}{3} (\bibinfo{year}{2021}), \bibinfo{pages}{537--555}.
\newblock


\bibitem[Shrivastava et~al\mbox{.}(2023)]%
        {repositorylevel}
\bibfield{author}{\bibinfo{person}{Disha Shrivastava}, \bibinfo{person}{Hugo Larochelle}, {and} \bibinfo{person}{Daniel Tarlow}.} \bibinfo{year}{2023}\natexlab{}.
\newblock \showarticletitle{Repository-level prompt generation for large language models of code}. In \bibinfo{booktitle}{\emph{International Conference on Machine Learning}}. PMLR, \bibinfo{pages}{31693--31715}.
\newblock


\bibitem[Stack Overflow({[n.\,d.]})]%
        {SO}
Stack Overflow \bibinfo{year}{[n.\,d.]}\natexlab{}.
\newblock
\newblock
\urldef\tempurl%
\url{https://stackoverflow.com/}
\showURL{%
\tempurl}


\bibitem[Tang et~al\mbox{.}(2024)]%
        {tang2024biocoder}
\bibfield{author}{\bibinfo{person}{Xiangru Tang}, \bibinfo{person}{Bill Qian}, \bibinfo{person}{Rick Gao}, \bibinfo{person}{Jiakang Chen}, \bibinfo{person}{Xinyun Chen}, {and} \bibinfo{person}{Mark~B Gerstein}.} \bibinfo{year}{2024}\natexlab{}.
\newblock \showarticletitle{BioCoder: a benchmark for bioinformatics code generation with large language models}.
\newblock \bibinfo{journal}{\emph{Bioinformatics}} \bibinfo{volume}{40}, \bibinfo{number}{Supplement\_1} (\bibinfo{year}{2024}), \bibinfo{pages}{i266--i276}.
\newblock


\bibitem[Tao et~al\mbox{.}(2024)]%
        {tao2024magis}
\bibfield{author}{\bibinfo{person}{Wei Tao}, \bibinfo{person}{Yucheng Zhou}, \bibinfo{person}{Yanlin Wang}, \bibinfo{person}{Wenqiang Zhang}, \bibinfo{person}{Hongyu Zhang}, {and} \bibinfo{person}{Yu Cheng}.} \bibinfo{year}{2024}\natexlab{}.
\newblock \showarticletitle{Magis: Llm-based multi-agent framework for github issue resolution}.
\newblock \bibinfo{journal}{\emph{arXiv preprint arXiv:2403.17927}} (\bibinfo{year}{2024}).
\newblock


\bibitem[TechCrunch({[n.\,d.]})]%
        {TechCrunch}
TechCrunch \bibinfo{year}{[n.\,d.]}\natexlab{}.
\newblock
\newblock
\urldef\tempurl%
\url{https://techcrunch.com/category/artificial-intelligence/}
\showURL{%
\tempurl}


\bibitem[Ugare et~al\mbox{.}(2024)]%
        {ugare2024improving}
\bibfield{author}{\bibinfo{person}{Shubham Ugare}, \bibinfo{person}{Tarun Suresh}, \bibinfo{person}{Hangoo Kang}, \bibinfo{person}{Sasa Misailovic}, {and} \bibinfo{person}{Gagandeep Singh}.} \bibinfo{year}{2024}\natexlab{}.
\newblock \showarticletitle{Improving llm code generation with grammar augmentation}.
\newblock \bibinfo{journal}{\emph{arXiv preprint arXiv:2403.01632}} (\bibinfo{year}{2024}).
\newblock


\bibitem[Wang et~al\mbox{.}(2024a)]%
        {wang2024sparsecoder}
\bibfield{author}{\bibinfo{person}{Yanlin Wang}, \bibinfo{person}{Yanxian Huang}, \bibinfo{person}{Daya Guo}, \bibinfo{person}{Hongyu Zhang}, {and} \bibinfo{person}{Zibin Zheng}.} \bibinfo{year}{2024}\natexlab{a}.
\newblock \showarticletitle{SparseCoder: Identifier-Aware Sparse Transformer for File-Level Code Summarization}.
\newblock \bibinfo{journal}{\emph{arXiv preprint arXiv:2401.14727}} (\bibinfo{year}{2024}).
\newblock


\bibitem[Wang et~al\mbox{.}(2024b)]%
        {wang2024beyond}
\bibfield{author}{\bibinfo{person}{Yanlin Wang}, \bibinfo{person}{Tianyue Jiang}, \bibinfo{person}{Mingwei Liu}, \bibinfo{person}{Jiachi Chen}, {and} \bibinfo{person}{Zibin Zheng}.} \bibinfo{year}{2024}\natexlab{b}.
\newblock \showarticletitle{Beyond functional correctness: Investigating coding style inconsistencies in large language models}.
\newblock \bibinfo{journal}{\emph{arXiv preprint arXiv:2407.00456}} (\bibinfo{year}{2024}).
\newblock


\bibitem[Wang et~al\mbox{.}(2024c)]%
        {wang2024rlcoder}
\bibfield{author}{\bibinfo{person}{Yanlin Wang}, \bibinfo{person}{Yanli Wang}, \bibinfo{person}{Daya Guo}, \bibinfo{person}{Jiachi Chen}, \bibinfo{person}{Ruikai Zhang}, \bibinfo{person}{Yuchi Ma}, {and} \bibinfo{person}{Zibin Zheng}.} \bibinfo{year}{2024}\natexlab{c}.
\newblock \showarticletitle{Rlcoder: Reinforcement learning for repository-level code completion}.
\newblock \bibinfo{journal}{\emph{arXiv preprint arXiv:2407.19487}} (\bibinfo{year}{2024}).
\newblock


\bibitem[Wei et~al\mbox{.}(2024)]%
        {magicoder}
\bibfield{author}{\bibinfo{person}{Yuxiang Wei}, \bibinfo{person}{Zhe Wang}, \bibinfo{person}{Jiawei Liu}, \bibinfo{person}{Yifeng Ding}, {and} \bibinfo{person}{Lingming Zhang}.} \bibinfo{year}{2024}\natexlab{}.
\newblock \showarticletitle{Magicoder: Empowering code generation with oss-instruct}. In \bibinfo{booktitle}{\emph{Forty-first International Conference on Machine Learning}}.
\newblock


\bibitem[Xia et~al\mbox{.}(2024)]%
        {evoeval}
\bibfield{author}{\bibinfo{person}{Chunqiu~Steven Xia}, \bibinfo{person}{Yinlin Deng}, {and} \bibinfo{person}{LINGMING ZHANG}.} \bibinfo{year}{2024}\natexlab{}.
\newblock \showarticletitle{Top Leaderboard Ranking = Top Coding Proficiency, Always? EvoEval: Evolving Coding Benchmarks via {LLM}}. In \bibinfo{booktitle}{\emph{First Conference on Language Modeling}}.
\newblock
\urldef\tempurl%
\url{https://openreview.net/forum?id=zZa7Ke7WAJ}
\showURL{%
\tempurl}


\bibitem[Xu et~al\mbox{.}(2022)]%
        {polycoder}
\bibfield{author}{\bibinfo{person}{Frank~F. Xu}, \bibinfo{person}{Uri Alon}, \bibinfo{person}{Graham Neubig}, {and} \bibinfo{person}{Vincent~J. Hellendoorn}.} \bibinfo{year}{2022}\natexlab{}.
\newblock \bibinfo{title}{A Systematic Evaluation of Large Language Models of Code}.
\newblock
\newblock
\showeprint[arxiv]{2202.13169}~[cs.PL]
\urldef\tempurl%
\url{https://arxiv.org/abs/2202.13169}
\showURL{%
\tempurl}


\bibitem[Yang et~al\mbox{.}(2024)]%
        {yang2024harnessing}
\bibfield{author}{\bibinfo{person}{Jingfeng Yang}, \bibinfo{person}{Hongye Jin}, \bibinfo{person}{Ruixiang Tang}, \bibinfo{person}{Xiaotian Han}, \bibinfo{person}{Qizhang Feng}, \bibinfo{person}{Haoming Jiang}, \bibinfo{person}{Shaochen Zhong}, \bibinfo{person}{Bing Yin}, {and} \bibinfo{person}{Xia Hu}.} \bibinfo{year}{2024}\natexlab{}.
\newblock \showarticletitle{Harnessing the power of llms in practice: A survey on chatgpt and beyond}.
\newblock \bibinfo{journal}{\emph{ACM Transactions on Knowledge Discovery from Data}} \bibinfo{volume}{18}, \bibinfo{number}{6} (\bibinfo{year}{2024}), \bibinfo{pages}{1--32}.
\newblock


\bibitem[Yu et~al\mbox{.}(2024)]%
        {codereval}
\bibfield{author}{\bibinfo{person}{Hao Yu}, \bibinfo{person}{Bo Shen}, \bibinfo{person}{Dezhi Ran}, \bibinfo{person}{Jiaxin Zhang}, \bibinfo{person}{Qi Zhang}, \bibinfo{person}{Yuchi Ma}, \bibinfo{person}{Guangtai Liang}, \bibinfo{person}{Ying Li}, \bibinfo{person}{Qianxiang Wang}, {and} \bibinfo{person}{Tao Xie}.} \bibinfo{year}{2024}\natexlab{}.
\newblock \showarticletitle{Codereval: A benchmark of pragmatic code generation with generative pre-trained models}. In \bibinfo{booktitle}{\emph{Proceedings of the 46th IEEE/ACM International Conference on Software Engineering}}. \bibinfo{pages}{1--12}.
\newblock


\bibitem[Zan et~al\mbox{.}(2023)]%
        {private}
\bibfield{author}{\bibinfo{person}{Daoguang Zan}, \bibinfo{person}{Bei Chen}, \bibinfo{person}{Yongshun Gong}, \bibinfo{person}{Junzhi Cao}, \bibinfo{person}{Fengji Zhang}, \bibinfo{person}{Bingchao Wu}, \bibinfo{person}{Bei Guan}, \bibinfo{person}{Yilong Yin}, {and} \bibinfo{person}{Yongji Wang}.} \bibinfo{year}{2023}\natexlab{}.
\newblock \showarticletitle{Private-library-oriented code generation with large language models}.
\newblock \bibinfo{journal}{\emph{arXiv preprint arXiv:2307.15370}} (\bibinfo{year}{2023}).
\newblock


\bibitem[Zan et~al\mbox{.}(2022a)]%
        {zan-etal-2022-language}
\bibfield{author}{\bibinfo{person}{Daoguang Zan}, \bibinfo{person}{Bei Chen}, \bibinfo{person}{Zeqi Lin}, \bibinfo{person}{Bei Guan}, \bibinfo{person}{Wang Yongji}, {and} \bibinfo{person}{Jian-Guang Lou}.} \bibinfo{year}{2022}\natexlab{a}.
\newblock \showarticletitle{When Language Model Meets Private Library}. In \bibinfo{booktitle}{\emph{Findings of the Association for Computational Linguistics: EMNLP 2022}}, \bibfield{editor}{\bibinfo{person}{Yoav Goldberg}, \bibinfo{person}{Zornitsa Kozareva}, {and} \bibinfo{person}{Yue Zhang}} (Eds.). \bibinfo{publisher}{Association for Computational Linguistics}, \bibinfo{address}{Abu Dhabi, United Arab Emirates}, \bibinfo{pages}{277--288}.
\newblock
\urldef\tempurl%
\url{https://doi.org/10.18653/v1/2022.findings-emnlp.21}
\showDOI{\tempurl}


\bibitem[Zan et~al\mbox{.}(2022b)]%
        {zan2022cert}
\bibfield{author}{\bibinfo{person}{Daoguang Zan}, \bibinfo{person}{Bei Chen}, \bibinfo{person}{Dejian Yang}, \bibinfo{person}{Zeqi Lin}, \bibinfo{person}{Minsu Kim}, \bibinfo{person}{Bei Guan}, \bibinfo{person}{Yongji Wang}, \bibinfo{person}{Weizhu Chen}, {and} \bibinfo{person}{Jian-Guang Lou}.} \bibinfo{year}{2022}\natexlab{b}.
\newblock \showarticletitle{CERT: continual pre-training on sketches for library-oriented code generation}.
\newblock \bibinfo{journal}{\emph{arXiv preprint arXiv:2206.06888}} (\bibinfo{year}{2022}).
\newblock


\bibitem[Zhang et~al\mbox{.}(2023)]%
        {repocoder}
\bibfield{author}{\bibinfo{person}{Fengji Zhang}, \bibinfo{person}{Bei Chen}, \bibinfo{person}{Yue Zhang}, \bibinfo{person}{Jacky Keung}, \bibinfo{person}{Jin Liu}, \bibinfo{person}{Daoguang Zan}, \bibinfo{person}{Yi Mao}, \bibinfo{person}{Jian-Guang Lou}, {and} \bibinfo{person}{Weizhu Chen}.} \bibinfo{year}{2023}\natexlab{}.
\newblock \showarticletitle{Repocoder: Repository-level code completion through iterative retrieval and generation}.
\newblock \bibinfo{journal}{\emph{arXiv preprint arXiv:2303.12570}} (\bibinfo{year}{2023}).
\newblock


\bibitem[Zhang et~al\mbox{.}(2024b)]%
        {naturalcodebench}
\bibfield{author}{\bibinfo{person}{Shudan Zhang}, \bibinfo{person}{Hanlin Zhao}, \bibinfo{person}{Xiao Liu}, \bibinfo{person}{Qinkai Zheng}, \bibinfo{person}{Zehan Qi}, \bibinfo{person}{Xiaotao Gu}, \bibinfo{person}{Xiaohan Zhang}, \bibinfo{person}{Yuxiao Dong}, {and} \bibinfo{person}{Jie Tang}.} \bibinfo{year}{2024}\natexlab{b}.
\newblock \showarticletitle{NaturalCodeBench: Examining Coding Performance Mismatch on HumanEval and Natural User Prompts}.
\newblock \bibinfo{journal}{\emph{arXiv preprint arXiv:2405.04520}} (\bibinfo{year}{2024}).
\newblock


\bibitem[Zhang et~al\mbox{.}(2024a)]%
        {zhang2024llm}
\bibfield{author}{\bibinfo{person}{Ziyao Zhang}, \bibinfo{person}{Yanlin Wang}, \bibinfo{person}{Chong Wang}, \bibinfo{person}{Jiachi Chen}, {and} \bibinfo{person}{Zibin Zheng}.} \bibinfo{year}{2024}\natexlab{a}.
\newblock \showarticletitle{Llm hallucinations in practical code generation: Phenomena, mechanism, and mitigation}.
\newblock \bibinfo{journal}{\emph{arXiv preprint arXiv:2409.20550}} (\bibinfo{year}{2024}).
\newblock


\bibitem[Zheng et~al\mbox{.}(2024)]%
        {zheng2024towards}
\bibfield{author}{\bibinfo{person}{Dewu Zheng}, \bibinfo{person}{Yanlin Wang}, \bibinfo{person}{Ensheng Shi}, \bibinfo{person}{Ruikai Zhang}, \bibinfo{person}{Yuchi Ma}, \bibinfo{person}{Hongyu Zhang}, {and} \bibinfo{person}{Zibin Zheng}.} \bibinfo{year}{2024}\natexlab{}.
\newblock \showarticletitle{Towards more realistic evaluation of LLM-based code generation: an experimental study and beyond}.
\newblock \bibinfo{journal}{\emph{arXiv preprint arXiv:2406.06918}} (\bibinfo{year}{2024}).
\newblock


\bibitem[Zheng et~al\mbox{.}(2025)]%
        {humanevo}
\bibfield{author}{\bibinfo{person}{Dewu Zheng}, \bibinfo{person}{Yanlin Wang}, \bibinfo{person}{Ensheng Shi}, \bibinfo{person}{Ruikai Zhang}, \bibinfo{person}{Yuchi Ma}, {and} \bibinfo{person}{Hongyu Zhangand~Zibin Zheng}.} \bibinfo{year}{2025}\natexlab{}.
\newblock \showarticletitle{HumanEvo: An Evolution-aware Benchmark for More Realistic Evaluation of Repository-level Code Generation}. In \bibinfo{booktitle}{\emph{The 47th International Conference on Software Engineering (ICSE 2025)}}.
\newblock
\urldef\tempurl%
\url{https://arxiv.org/abs/2406.06918}
\showURL{%
\tempurl}


\bibitem[Zheng et~al\mbox{.}(2023c)]%
        {zheng2023codegeex}
\bibfield{author}{\bibinfo{person}{Qinkai Zheng}, \bibinfo{person}{Xiao Xia}, \bibinfo{person}{Xu Zou}, \bibinfo{person}{Yuxiao Dong}, \bibinfo{person}{Shan Wang}, \bibinfo{person}{Yufei Xue}, \bibinfo{person}{Lei Shen}, \bibinfo{person}{Zihan Wang}, \bibinfo{person}{Andi Wang}, \bibinfo{person}{Yang Li}, {et~al\mbox{.}}} \bibinfo{year}{2023}\natexlab{c}.
\newblock \showarticletitle{Codegeex: A pre-trained model for code generation with multilingual benchmarking on humaneval-x}. In \bibinfo{booktitle}{\emph{Proceedings of the 29th ACM SIGKDD Conference on Knowledge Discovery and Data Mining}}. \bibinfo{pages}{5673--5684}.
\newblock


\bibitem[Zheng et~al\mbox{.}(2023a)]%
        {survey2}
\bibfield{author}{\bibinfo{person}{Zibin Zheng}, \bibinfo{person}{Kaiwen Ning}, \bibinfo{person}{Jiachi Chen}, \bibinfo{person}{Yanlin Wang}, \bibinfo{person}{Wenqing Chen}, \bibinfo{person}{Lianghong Guo}, {and} \bibinfo{person}{Weicheng Wang}.} \bibinfo{year}{2023}\natexlab{a}.
\newblock \showarticletitle{Towards an understanding of large language models in software engineering tasks}.
\newblock \bibinfo{journal}{\emph{arXiv preprint arXiv:2308.11396}} (\bibinfo{year}{2023}).
\newblock


\bibitem[Zheng et~al\mbox{.}(2023b)]%
        {survey1}
\bibfield{author}{\bibinfo{person}{Zibin Zheng}, \bibinfo{person}{Kaiwen Ning}, \bibinfo{person}{Yanlin Wang}, \bibinfo{person}{Jingwen Zhang}, \bibinfo{person}{Dewu Zheng}, \bibinfo{person}{Mingxi Ye}, {and} \bibinfo{person}{Jiachi Chen}.} \bibinfo{year}{2023}\natexlab{b}.
\newblock \showarticletitle{A survey of large language models for code: Evolution, benchmarking, and future trends}.
\newblock \bibinfo{journal}{\emph{arXiv preprint arXiv:2311.10372}} (\bibinfo{year}{2023}).
\newblock


\bibitem[Zhong et~al\mbox{.}(2024)]%
        {zhong2024memorybank}
\bibfield{author}{\bibinfo{person}{Wanjun Zhong}, \bibinfo{person}{Lianghong Guo}, \bibinfo{person}{Qiqi Gao}, \bibinfo{person}{He Ye}, {and} \bibinfo{person}{Yanlin Wang}.} \bibinfo{year}{2024}\natexlab{}.
\newblock \showarticletitle{Memorybank: Enhancing large language models with long-term memory}. In \bibinfo{booktitle}{\emph{Proceedings of the AAAI Conference on Artificial Intelligence}}, Vol.~\bibinfo{volume}{38}. \bibinfo{pages}{19724--19731}.
\newblock


\bibitem[Zhu et~al\mbox{.}(2024a)]%
        {domaineval}
\bibfield{author}{\bibinfo{person}{Qiming Zhu}, \bibinfo{person}{Jialun Cao}, \bibinfo{person}{Yaojie Lu}, \bibinfo{person}{Hongyu Lin}, \bibinfo{person}{Xianpei Han}, \bibinfo{person}{Le Sun}, {and} \bibinfo{person}{Shing-Chi Cheung}.} \bibinfo{year}{2024}\natexlab{a}.
\newblock \bibinfo{title}{DOMAINEVAL: An Auto-Constructed Benchmark for Multi-Domain Code Generation}.
\newblock
\newblock
\showeprint[arxiv]{2408.13204}~[cs.AI]
\urldef\tempurl%
\url{https://arxiv.org/abs/2408.13204}
\showURL{%
\tempurl}


\bibitem[Zhu et~al\mbox{.}(2024b)]%
        {deepseek}
\bibfield{author}{\bibinfo{person}{Qihao Zhu}, \bibinfo{person}{Daya Guo}, \bibinfo{person}{Zhihong Shao}, \bibinfo{person}{Dejian Yang}, \bibinfo{person}{Peiyi Wang}, \bibinfo{person}{Runxin Xu}, \bibinfo{person}{Y Wu}, \bibinfo{person}{Yukun Li}, \bibinfo{person}{Huazuo Gao}, \bibinfo{person}{Shirong Ma}, {et~al\mbox{.}}} \bibinfo{year}{2024}\natexlab{b}.
\newblock \showarticletitle{DeepSeek-Coder-V2: Breaking the Barrier of Closed-Source Models in Code Intelligence}.
\newblock \bibinfo{journal}{\emph{arXiv preprint arXiv:2406.11931}} (\bibinfo{year}{2024}).
\newblock


\bibitem[Zhuo et~al\mbox{.}(2024)]%
        {zhuo2024bigcodebenchbenchmarkingcodegeneration}
\bibfield{author}{\bibinfo{person}{Terry~Yue Zhuo}, \bibinfo{person}{Minh~Chien Vu}, \bibinfo{person}{Jenny Chim}, \bibinfo{person}{Han Hu}, \bibinfo{person}{Wenhao Yu}, \bibinfo{person}{Ratnadira Widyasari}, \bibinfo{person}{Imam Nur~Bani Yusuf}, \bibinfo{person}{Haolan Zhan}, \bibinfo{person}{Junda He}, \bibinfo{person}{Indraneil Paul}, \bibinfo{person}{Simon Brunner}, \bibinfo{person}{Chen Gong}, \bibinfo{person}{Thong Hoang}, \bibinfo{person}{Armel~Randy Zebaze}, \bibinfo{person}{Xiaoheng Hong}, \bibinfo{person}{Wen-Ding Li}, \bibinfo{person}{Jean Kaddour}, \bibinfo{person}{Ming Xu}, \bibinfo{person}{Zhihan Zhang}, \bibinfo{person}{Prateek Yadav}, \bibinfo{person}{Naman Jain}, \bibinfo{person}{Alex Gu}, \bibinfo{person}{Zhoujun Cheng}, \bibinfo{person}{Jiawei Liu}, \bibinfo{person}{Qian Liu}, \bibinfo{person}{Zijian Wang}, \bibinfo{person}{David Lo}, \bibinfo{person}{Binyuan Hui}, \bibinfo{person}{Niklas Muennighoff}, \bibinfo{person}{Daniel Fried}, \bibinfo{person}{Xiaoning Du}, \bibinfo{person}{Harm de
  Vries}, {and} \bibinfo{person}{Leandro~Von Werra}.} \bibinfo{year}{2024}\natexlab{}.
\newblock \bibinfo{title}{BigCodeBench: Benchmarking Code Generation with Diverse Function Calls and Complex Instructions}.
\newblock
\newblock
\showeprint[arxiv]{2406.15877}~[cs.SE]
\urldef\tempurl%
\url{https://arxiv.org/abs/2406.15877}
\showURL{%
\tempurl}


\end{thebibliography}

\end{document}